\documentclass[12pt]{iopart}
\usepackage{graphicx}
\usepackage{amssymb}
\usepackage{ulem}
\usepackage{subfig}
\newcommand{\comma}{,~}
\usepackage{feynmf}
\newcommand{\uul}[1]{\uuline{#1}{}}
\newcommand{\eqref}[1]{(\ref{#1})}
\renewcommand{\k}{\underline{k}}

\begin{document}

\title[Conserving approximations in direct perturbation theory]{Conserving approximations in direct perturbation theory: new semianalytical impurity solvers and their application to general lattice problems}
\author{Norbert Grewe$^1$, Sebastian Schmitt$^1$, Torben Jabben$^1$ and  Frithjof B. Anders$^2$}

\address{$^1$Institut f\"{u}r Festk\"{o}rperphysik, Technische Universit\"{a}t Darmstadt, 
Hochschulstr. 6, D-64289 Darmstadt, Germany}
\address{$^2$  Institut f\"{u}r Theoretische Physik, Universit\"{a}t Bremen, 
  P.O.Box 330 440, D-28334 Bremen, Germany}

\begin{abstract}
For the treatment of interacting electrons in crystal lattices
approximations based on the picture of effective sites, coupled in a
self-consistent fashion, have proven very useful. Particularly in the
presence of strong local correlations, a local approach to the
problem, combining a powerful method for the short ranged
interactions with the lattice propagation part of the dynamics,
determines the quality of results to a large extent. For a
considerable time the non crossing approximation (NCA) in direct
perturbation theory, an approach originally developed by Keiter for
the Anderson impurity model, built a standard for the description of
the local dynamics of interacting electrons. In the last couple of
years exact methods like the numerical renormalization group (NRG)
as pioneered by Wilson, have surpassed this approximation as
regarding the description of the low energy regime. We present an
improved approximation level of direct perturbation theory
for  finite Coulomb repulsion $U$, 
the crossing approximation one (CA1) and discuss its connections with
other generalizations of NCA. 
CA1 incorporates 
all processes up to fourth order in the hybridization strength $V$
in a self-consistent skeleton expansion, retaining
the full energy dependence of the vertex functions.
We reconstruct the local approach to
the lattice problem from the point of view of cumulant perturbation
theory in a very general way and discuss the proper use of impurity
solvers for this purpose. Their reliability can be tested in
applications to e.g.\ the Hubbard model and the Anderson-lattice
model. We point out shortcomings of existing impurity solvers and
improvements gained with CA1 in this context.\\[2mm]

{\bf This paper is dedicated to the memory of Hellmut Keiter.}
\end{abstract}



\maketitle

\tableofcontents

\section{Introduction}
 In a key paper~\cite{Quelle1}, H. Keiter and J. C. Kimball in
1971 described a new perturbational method for treating the problem
of an impurity with strong local Coulomb matrix elements, embedded
in a metallic host. Their guideline was to preserve the local
correlations from the outset, contrary to Hartree-Fock theory or to
other decoupling schemes, and to keep the interpretation of
individual contributions as physical processes. The particular
difficulties to be surmounted arose from the fact, that choosing
hybridization or intersite transfer of single particles as the
perturbation leaves an interacting local shell as the unperturbed
part of the Hamiltonian. In such a case the well known machinery of
Feynman diagrammatics cannot be used, including Wicks theorem and
linked cluster expansions. The solution found used time-ordered
pieces of Feynman processes, visualized as Goldstone diagrams, and
organized them in the form of Brillouin-Wigner perturbation theory
with real energy variables. In this early formulation of the theory
the need to regularize vanishing energy denominators prevented
extensive studies to infinite perturbational orders, which are
necessary in the presence of infrared divergencies, encountered e.g.\
in the Kondo problem. These can be thought of as to arise from
degeneracies in a classical part of the Hamiltonian; they are then
lifted by quantum fluctuations. Early applications of the technique
can be found in ~\cite{Quelle2}, where leading logarithmically
divergent terms are summed to all orders to generate finite results,
and in ~\cite{Quelle3,Quelle4}, where also generalizations of
the formalism to more general local shell structures and to the
genuine lattice problem were discussed.

The technique of Keiter and Kimball originally was designed in
connection with the Kondo problem. A revival occurred around 1980,
when metallic compounds exhibiting the intermediate valence
phenomenon stayed in the focus of experimentalists~\cite{Quelle5},
and somewhat later, when the existence of very heavy quasiparticles
(Heavy Fermions) was revealed in such compounds containing ions with
active 4f- or 5f-shells~\cite{Quelle6}. In particular the discovery
of Heavy Fermion superconductors [7] spurred the investigation of
lattice models with strong local correlations. A breakthrough in
theory occurred in 1983, when it was learnt how to handle direct
perturbation theory completely in the complex plane thus
circumventing the regularization problem. The first independent
extensive studies~\cite{Quelle8,Quelle9} of impurity problems were
based on the leading skeleton diagram contributions to the dynamics
of ionic shell states, requiring the solution of self consistently
coupled singular integral equations. The contributing diagrams are
completely characterized as not containing crossing band electron
lines. This approximation (NCA) was shown
to adequately describe dynamical properties like excitation spectra
in addition to thermodynamic quantities, the latter of which being
known from Wilson's implementation of the numerical renormalization
group~\cite{Quelle10}. NCA allows to study~\cite{Quelle8} the
temperature dependent formation of the Abrikosov-Suhl resonance
(ASR)~\cite{Quelle11} with an assessment of the systematic
shortcomings via a comparison to the resonant level model, the limit
of vanishing spin degeneracy. NCA can also be characterized as a
conserving approximation in the sense of Baym and
Kadanoff~\cite{Quelle12} and can consistently be applied to
calculate various properties of the strongly correlated impurity
problem~\cite{Quelle9}. Also in 1983, a reformulation of
perturbation theory was presented~\cite{Quelle13}, which allowed for
the use of Feynman diagrammatics via the introduction of auxiliary
particles (Slave Bosons). Correlated local states are reintroduced
in this approach by a constraint on the larger Hilbert space and a
corresponding projection onto the physical sector after
resummations. The Slave Boson method stands in one to one
correspondence to the formulation via direct  perturbation theory and thus
contributes to the same line of development.

Merits and shortcomings of the NCA for the Anderson impurity with
infinite local Coulomb repulsion meanwhile are well known, for
example as result of the early numerical
studies~\cite{Quelle8,Quelle14} or of the exact analysis of the case
with a flat electron-hole symmetric conduction band at zero
temperature~\cite{Quelle15}. NCA e.g.\ captures the exponential part
of the Kondo temperature $T_K$, the dynamically generated energy
scale below which a local Fermi liquid is formed due to spin
compensation, but not the prefactor. It does not furnish the correct
values of the threshold exponents, connected with the time
development of ionic states as known from the X-ray absorption
problem~\cite{Quelle16,Quelle17}. Its accuracy increases with
increasing (orbital) degeneracy of the ionic level, and NCA may even
become a fully acceptable approximation for a multi-channel
situation~\cite{Quelle18}.

The need for improved treatments of the
impurity problem with correlated electrons, however, turned out to
be even more important, when it became a building block in theories
of lattice systems ~\cite{Quelle19,Quelle20,Quelle21}, using the
concept of effective sites. It turned out that e.g.\ the coherence
forming in the low temperature regime of lattices can only be
correctly retrieved by a proper incorporation of the effective local
Fermi liquid. The necessary increase in the quality of results for
lattice models was largely driven by the use of numerically exact
methods as impurity solvers, i.e.\ quantum Monte Carlo (QMC) and
numerical renormalization group (NRG)~\cite{Quelle22}. NRG in
particular, in its extension of the original static version to
dynamical quantities~\cite{Quelle23}, has been developed to a useful
and convenient tool for a description of the low energy
region~\cite{Quelle24}. When combined with the (cluster-) dynamical
mean field theory (DMFT) it opens the perspective for a proper
description of ground states and correlations in the lattice.

Still there remain problems: (1) Non local matrix
elements of the Coulomb interaction can be strong, too, and may
enforce a nonlocal approach with links between sites the outset, 
as e.g.\ in pyrochlore lattices, where for
certain fillings classical ground states characterized by the
tetrahedron rule build the arena for quantum
fluctuations~\cite{Quelle25}. (2) The use of clusters
as larger local building blocks in a lattice theory greatly enlarges
the local space to be diagonalized in the
beginning~\cite{Quelle26,Quelle27} and enforces a restricted choice
of states and/or the use of simpler impurity solvers like e.g.\ a
simplified finite U-version of NCA (SNCA), especially if a
self consistent determination of one-particle states and matrix
elements is aimed at as part of the solution of a lattice
problem~\cite{Quelle28}. (3) The numerical methods
mentioned do not work equally well in different energy regimes.

NRG, for example, often leads to a poor description at high excitation
energies. Depending on tunable parameters in lattice models
corresponding to temperature, pressure, chemical composition,
doping, etc. a wealth of phases and corresponding transitions is
found; even a transmutation of underlying pictures or concepts may
occur, as e.g.\ from magnetism of (nearly) stable local magnetic
moments coupled via short ranged exchange interactions to itinerant
forms of magnetism to be described by an effective Stoner-theory for
bands of itinerant quasiparticles~\cite{Quelle21,Quelle29}. In these
cases a good description of a large regime of excitation energies is
desired, which requires impurity solvers, which work equally well in
a broad range of energies and allow for a controlled approach to
known limiting cases. Facing these difficulties for a more complete
understanding and description of lattice systems, further
developments of methods with analytical background seem necessary.

\medskip
In this paper we will present a version of direct perturbation
theory for an (effective) impurity, which combines elements of some
existing improvements of NCA and goes beyond them in some respects.
An essential ingredient of this new stage of approximation are
processes with crossing band electron lines, as depicted along a
(imaginary) time axis, and hence the name crossing approximation one
(CA1). In the following section 2 we will give a very short account
of direct perturbation theory, describe CA1 and comment on its
relation to NCA and its extensions as accounted for in the
literature so far. Section 3 contains some calculated spectra for
the Anderson impurity model and a comparison to former
approximations and, most important, to results obtained with the
NRG. As it will become apparent, CA1 turns out to be a rather good
impurity solver in the whole range of energies and can be applied to
calculate excitation spectra, ionic propagators and
susceptibilities. A conceptual bridge between impurity and lattice
theories is outlined in section 4, which allows to recover the
results of lattice theories like XNCA~\cite{Quelle20} and
DMFT~\cite{Quelle22,Quelle30} in a general fashion.

 As an example
for the usefulness of CA1 in this context a calculation for the
Anderson lattice is presented. The concluding section 5 contains
remarks about an application of direct perturbation theory to
susceptibilities and magnetic phases of lattice models and about
improvements regarding nonlocal correlations as well as a new
impurity solver, i.e.\ a CA2-project.

\section{Description of CA1}
\subsection{Introductory remarks on  direct perturbation theory}
A typical setup for the application of direct perturbation theory
uses a Hamiltonian $H=H_{0\ell}+H_{0c}+V$ with the following parts:
$H_{0\ell}\equiv{H_{0\ell}}(\{f_{m\sigma},f^+_{m\sigma}\})$ contains
the dynamics of interacting electrons in local one-particle states
with quantum numbers m and $\sigma$ (pseudo spin) and is expressed
via corresponding annihilation (creation) operators
$f_{m\sigma}^{(+)}.\:H_{0c}\equiv{H_{0c}}(\{c_{\underline{k}\sigma},c^+_{\underline{k}\sigma}\})$
describes a reservoir of noninteracting electrons in Bloch states (a
band index is suppressed here), and
$V=V(\{f_{m\sigma},f^+_{m\sigma},c_{\underline{k}\sigma},c^+_{\underline{k}\sigma}\})$
is a hybridization or transfer between local and band states, which
is likewise expressed via elementary one-particle processes.
$H_{0\ell}$ acts on a local Fock-space of finite dimension, typically
one or a few valence shells or orbitals, and can in principle be
diagonalized. A basis of eigenstates $\mid{n_0,{M}}\:\rangle$
("ionic states") is denominated by a local particle number $n_0$ and
$a$ set of many-body quantum numbers $M$ specifying angular momenta
or crystal field levels. With the operators
$X_{{n'_0}M',n_0M}\equiv\mid{n'_0}M'\rangle\langle{n}_0{M}\mid$ and
corresponding $n_0$-particle energies $E_{n_o{M}}$ the local
Hamiltonian reads
\begin{eqnarray}
H_{0\ell}=\sum\limits_{{n_0},M}E_{n_0 M}X_{{n_0}M,{n_0}M}
\label{Gleichung1}
\end{eqnarray}
where only projectors onto the eigenstates appear. The terms "local"
or  "ionic" do not necessarily imply one single atom. The formalism
equally well applies to local subsystems of molecular type or to
local clusters. A transcription of $V$ to local many body states
involves via
\begin{eqnarray}
f^{(+)}_{m\sigma}=\sum\limits_{n_0}\sum\limits_{M,M'}\alpha^{(*)}_{m\sigma}(n_0-1M',n_0M)X^{(+)}_{n_0-1M',n_0M}
\label{Gleichung2}
\end{eqnarray}
the set of ionic transfer operators $X_{{n'_0}M',{n_0}M}$ with
$n'_0=n_0\pm1$. Using  $V$ as the perturbation, processes of direct
perturbation theory are constructed from elementary absorption or
emission events of band electrons from a local shell state at fixed
(imaginary) times with amplitudes given by the coefficients
$\alpha_{m\sigma}(n'_0M',n_0M)$ in
(\ref{Gleichung2})~\cite{Quelle3}. Insofar it can be expected that
e.g.\ the partition function can be cast into a form, where the
contribution of each particular ionic state becomes apparent:
\begin{eqnarray}
Z&=Tre^{-\beta
H}=\oint\limits_\mathcal{C}\:\frac{dz}{2\pi{i}}\:e^{-{{\beta}z}}Tr_{\ell}Tr_{c}(z-H)^{-1}
=Z_{0c}\sum\limits_{{n_0},M}\oint\limits_\mathcal{C}\:
\frac{dz}{2\pi{i}}\:e^{-{{\beta}z}}P_{n_0{M}}(z)\nonumber\\
&=Z_{0c}\sum\limits_{n_0,{M}}\int{d\omega{e}}^{-{{\beta}\omega}}\varrho_{n_0{M}}(\omega).
\label{Gleichung3}
\end{eqnarray}
$Z_{0c}=Tr_c\:e^{-{\beta}H_{0c}}$ is the partition function for the
band part alone and
$\varrho_{nM_0}(\omega)=-\frac{1}{\pi}Im\:P_{n_0{M}}(\omega+i\delta)$
is the spectral intensity of the ionic state $|{n}_0{M}\rangle$,
which evolves with the propagator $P_{n_0M}(z)$.
 A straightforward concept of
irreducibility with respect to intermediate ionic states allows for
the introduction of irreducible ionic selfenergies, with analytical
properties as usual,
\begin{eqnarray}
P_{n_0{M}}(z)=\left(z-E_{n_0{M}}-\Sigma_{n_0{M}}(z)\right)^{-1},
\label{Gleichung6}
\end{eqnarray}
and a corresponding perturbation expansion. The processes
contributing to these selfenergies $\Sigma_{n_0{M}}(z)$ will in the
following be constructed from skeleton diagrams, so that the ionic
propagators $P_{n_0{M}}(z)$ are to be determined selfconsistently
from a set of coupled integral equations.

Representations for general Greensfunctions, which in correspondence
with the partition function (\ref{Gleichung3}) are expressed as
convolutions of ionic propagators, can also be derived along the
lines sketched above. We consider in particular the local
one-particle Greensfunction,
\begin{eqnarray}
F_{m\sigma}(\tau)&=-\langle{T}(f_{m\sigma}(\tau)f^+_{m\sigma})\rangle\nonumber\\
&=-\sum_{n_0,\widetilde{n}_{0}}\sum_{{M_1},{M'_1}}\sum_{{M_2},{M'_2}}\alpha_{m\sigma}(n_{0}-1{M'_1},n_{0}M_1)
\alpha^*_{m\sigma}(\widetilde{n}_0{-1}{M'_2},\widetilde{n}_0{M_2})\nonumber\\
&\qquad\langle{T}(X_{n_{0}-1{M'_1},n_0{M_1}}(\tau)X_{\widetilde{n}_0{M_2},\widetilde{n}_0{-1}{M'_2}})\rangle,
\label{Gleichung7}
\end{eqnarray}
the Fourier-coefficients of which at Matsubara frequencies
$\omega_n=\frac{(2n+1)\pi}{\beta}\:(n\in\mathbb{Z})$ give, after
analytical continuation $i\omega_n\rightarrow\omega+i\delta$ to the
upper border of the real frequency axis, the local one particle
excitation spectrum
$\varrho_{m\sigma}(\omega)=-\frac{1}{\pi}Im{F_{m\sigma}}(\omega+i\delta)$.
The complete setup of nonstandard direct perturbation theory is well documented,
including the diagrammatic rules for the processes to be discussed in 
the following~\cite{Quelle3,Quelle8,Quelle31}.

\medskip

\subsection{General features of crossing and non crossing  approximations for the SIAM}
First comprehensive studies of direct perturbation theory for the Anderson 
impurity model (SIAM)
\begin{equation}
  \label{eq:SIAMH}
  \hat H =\sum_\sigma \left(
    \epsilon_\ell \:\hat{f}^\dagger_{\sigma}\hat{f}_{\sigma}
    +\frac{U}{2}\, \hat{n}^f_{\sigma}\hat{n}^f_{\bar{\sigma}}
  \right)
  +\sum_{\k,\sigma}\epsilon_{\k}\,\hat{c}^\dagger_{\k\sigma}\hat{c}_{\k\sigma}
  +\frac{1}{\sqrt{N}}\sum_{\k,\sigma} \left(V_{\k}\,\hat{c}^\dagger_{\k\sigma}\hat{f}_{\sigma}+h.c. 
  \right)
\end{equation}
concentrated on the limit $U=\infty$ of infinite local Coulomb
repulsion and were termed NCA~\cite{Quelle8,Quelle9,Quelle32}. 
They were based on the
leading skeletons of order $V^2$ to the ionic self energies
$\Sigma_0(z)$ and $\Sigma_{1\sigma}(z)$ and furnished a
qualitatively correct picture e.g.\ for the temperature dependent
formation of the Abrikosov-Suhl resonance (ASR), the most prominent
many-body signature of the Kondo-effect~\cite{Quelle33} in the
local one-particle spectrum of the model. The particular aspect of
degeneracy $\nu$ of the local level, i.e.\ $\nu=2$ for the two
possible $z-$components of spin in the original SIAM but higher
$\nu$ as in Ce-compounds with $\nu=6$ becoming possible through
orbital degeneracy, drew much attention: In the limit
$\nu\rightarrow\infty$, using a proper scaling
$V\rightarrow\frac{V}{\sqrt{\nu}}$, the NCA-results become
increasingly valid~\cite{Quelle2,Quelle13,Quelle34}, and for
$\nu\rightarrow1$ on the other hand a trivially solvable resonant
level model emerges. Whereas the limit of large $\nu$ gave reason
for classification schemes of diagrams in orders of $\frac{1}{\nu}$,
the limit $\nu=1$ arose hopes of reconstructing an exact solution of
a simple model by direct perturbation theory thus completely
clarifying the systematics of diagrams for all cases of $\nu$.

This hope was not fulfilled up to now, although Keiter presented an exact
solution of SIAM for the zero-bandwidth limit unravelling the full
diagrammatics for this simpler case~\cite{Quelle35}. It became clear
then, that progress with the direct perturbation approach had to be
worked out stepwise by including more important classes of skeleton
diagrams into the calculations.

In the following we discuss the
systematics of these approximations for the SIAM by concentrating on
the vertices, which allows for writing down several quantities in a
compact and rigorous form, i.e.\ the ionic selfenergies
\begin{eqnarray}
\Sigma_0(z)&=\sum\limits_\sigma\int{dx}\:D_\sigma(x)f(x)\Lambda_{0,1\sigma}(z,x)P_{1\sigma}(z+x),\nonumber\\
\Sigma_{1\sigma}(z)&=\int{dx}f(x)\Big[D_\sigma(-x)\Lambda_{0,1\sigma}(z+x,-x)P_0(z+x)
\label{Gleichung11}
\\\nonumber &
\phantom{=\int{dx}f(x)\Big[}
+D_{-\sigma}(x)
\Lambda_{2,1\sigma}(z+x,-x)P_2(z+x)\Big],\nonumber\\
\Sigma_2(z)&=\sum\limits_\sigma\int{dx}D_{-\sigma}(-x)f(x)\Lambda_{2,1\sigma}(z,x)P_{1\sigma}(z+x),
\nonumber
\end{eqnarray}
and the local one-particle Greensfunction~\eqref{Gleichung7}, which
in the special case of the SIAM contains only two contributions:
\begin{eqnarray}\label{Gleichung12}
  F_{\sigma}(z)&=\frac{1}{Z_\ell}\oint\limits_\mathcal{C}\:\frac{dz'}{2\pi{i}}e^{-\beta{z'}}
\Big[\Lambda_{0,1\sigma}(z',z)P_0(z')P_{1\sigma}(z+z')
\\\nonumber
& \phantom{=\frac{1}{Z_\ell}\oint\limits_\mathcal{C}\:\frac{dz'}{2\pi{i}}e^{-\beta{z'}}\Big[}
+\Lambda_{2,1-\sigma}(z',-z)P_2(z')P_{1-\sigma}(z-z')
\Big].
\end{eqnarray}
Here we have introduced the hybridization intensity
$D_\sigma(\epsilon)= \frac{V(\epsilon)^2}{N}\:
\sum_{\underline{k}}\delta(\epsilon-\epsilon_{\underline{k}\sigma})=V(\epsilon)^2
\varrho_{0c}(\epsilon)$ and the (perturbed) local partition function
$Z_\ell=\frac{Z}{Z_{0c}}$. Expressions for higher Greens functions
take an analogous form; we only add here a formula for the dynamical
magnetic susceptibility (leaving out prefactors
$\left(\frac{1}{2}g\mu_\beta\right)^2$),
\begin{eqnarray}
\chi_{mag}(z)=-\frac{1}{Z_\ell}\oint\limits_\mathcal{C}\frac{dz'}{2\pi{i}}\:e^{-\beta{z'}}\sum\limits_\sigma
\Lambda_{\sigma,\sigma}(z',z)P_{1\sigma}(z')P_{1\sigma}(z+z'),
\label{Gleichung13}
\end{eqnarray}
which involves a separate kind of vertex $\Lambda_{\sigma,\sigma}$.
Eqs.~(\ref{Gleichung11}) to (\ref{Gleichung13}) are graphically
represented in figure~\ref{Figure1}; observe identities like
$\Lambda_{0,1\sigma}(z-z',z')=\Lambda_{1\sigma,0}(z,z')$ for setting
up the equations from there. 

In the SNCA, which can be viewed as the
simplest nontrivial approximation for all values of $U$, the vertex
functions are all taken without any vertex corrections:
\begin{eqnarray}
\textrm{SNCA:}
\quad\Lambda_{0,1\sigma}=\Lambda_{2,1\sigma}=\Lambda_{\sigma,\sigma}\equiv{1}\:.
\label{Gleichung14}
\end{eqnarray}
\begin{fmffile}{fmf_nca1}
\begin{fmfshrink}{0.8}
\fmfcmd{%
    style_def wiggly_arrow expr p =
     cdraw (wiggly p);
     cfill (arrow p)
    enddef;
    style_def dbl_wiggly_arrow expr p =
     draw_double (wiggly p) ;
     cfill (arrow p);
    enddef;
}
\begin{figure}[t]
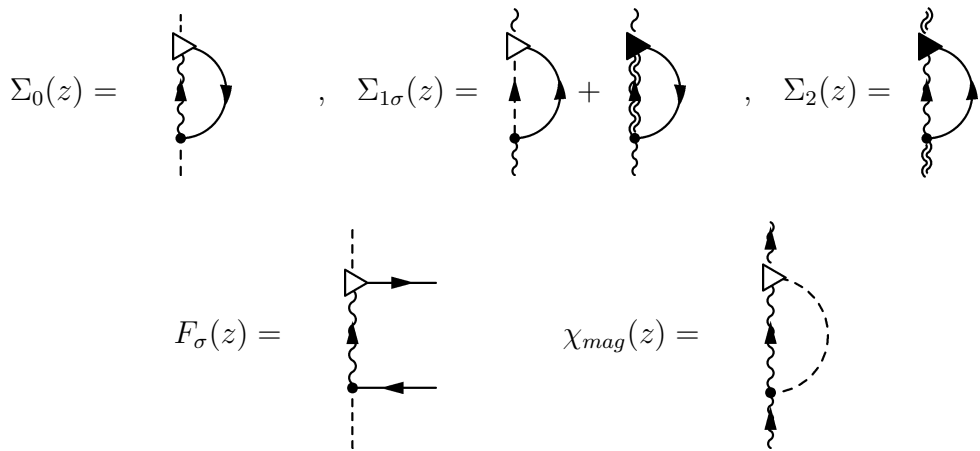

  \begin{center}
    \input{sig0_nca.tex}\\[6mm]
    \input{gf_nca.tex}%
    \hspace*{2cm}%
    \input{chi_nca.tex}
  \end{center}
  \caption{Diagrammatic representation of ionic selfenergies, local
    one-particle Greenfunction and magnetic susceptibility for a
    single-impurity Anderson model (SIAM) in direct perturbation theory.
    Physical processes are arranged vertically along an imaginary time
    axis (broken line) which bears an energy variable z after
    Laplace-transformation. Presence of an electron in the local shell
    is indicated via a wiggly line on this time axis. Excitations of
    band electrons (straight lines) take place at hybridization vertices
    (dots on the time axis). Due to time-rotational invariance all
    vertex corrections in these diagrams can be collected at one of the
    vertices, which is drawn as a triangle.}
  \label{Figure1}
\end{figure}
\end{fmfshrink}
\end{fmffile}
The original NCA constitutes the $U\rightarrow\infty$-limit hereof and is
obtained by ignoring the doubly occupied state, i.e.\ by setting
$P_2\equiv0$.

\medskip

The NCA, and in consequence the SNCA for general values of $U$, can
only furnish qualitative insight into the dynamics of the SIAM,
since it is plagued by shortcomings. These are revealed in the
following ways: (1) Comparison with the
resonant-level limit $\nu\rightarrow1$ ends in an insufficient fit
to the virtual scattering resonance in the excitation spectrum, in
particular when the intermediate valence regime is approached
~\cite{Quelle8}. (2) Form and position of the ASR are
not in accord with the Friedel sum rule, most important for $\nu=2$,
and correspondingly the local self energy
$\widetilde{\Sigma}_\sigma(z)=\Sigma_\sigma(z)+i\Delta_A=z-\epsilon_\ell+i\Delta_A-F_{\sigma}(z)^{-1}$
does not comply with local Fermi-liquid properties~\cite{Quelle36}
$\left(\Delta_A=\pi{V}(0)^2\varrho_{0c}(0)\right.$ is the Anderson
width and $\epsilon_\ell=E_{1\sigma}-E_0$ the local one-particle
level). (3) threshold exponents, as taken from
the ionic propagators (see below) with values $\alpha_0=\frac{1}{3}$
and $\alpha_{1\sigma}=\frac{2}{3}$ in the NCA, do not agree with the
values known from the X-ray-absorption problem~\cite{Quelle17}.
(4) An exact analytical solution of the NCA-version
of Eqs.~(\ref{Gleichung11}) for zero temperature and a flat
conduction band density of states symmetric around the Fermi energy
reveals spurious features near the ASR ~\cite{Quelle15}, namely a
sharp spike showing up at the Fermi level below a ("pathology-")
temperature $T_p$, being still lower than $T_K$ in the Kondo regime.

Point (3) deserves some further comments, because it
hints to the particular singular structure of the ionic propagators
$P_{n_0M}(z)$, which causes difficulties in the numerical solution of the
system (\ref{Gleichung11}) of integral equations and also in
subsequent procedures like (\ref{Gleichung12}), (\ref{Gleichung13})
involving convolutions of several of the $P_{n_0M}$. As explained e.g.\ 
in~\cite{Quelle17}, these propagators develop a common threshold
at an energy $\omega=E_g<\epsilon_\ell$ for zero temperature, due to
a slow algebraic decay
$P_{n_0M}(t)\sim{e^{-\frac{i}{\hbar}E_g{t}}}/t^{\alpha_M}$ in the time
domain, i.e.\
\begin{equation}
\mathrm{Im}\:P_{n_0M}(\omega-i\delta)\sim1/(\omega-E_g)^{1-\alpha_M},\quad\alpha_0=\frac{n^2_\ell}{\nu},\quad
\alpha_{1\sigma}=1-2\frac{n_\ell}{\nu}+\frac{n^2_\ell}{\nu},\quad
\label{Gleichung15}
\end{equation}
$0<{n_\ell}\leq1$ being the occupation of the local level in the
Kondo-regime $0<\Delta_A<-\epsilon_\ell<{U}.$ At $T=0,\:{E_g}$ is
the lower endpoint of a branch cut in the functions $P_{n_0M}(z)$ along
the real axis $z=\omega>{E_g}$; it is this particular divergent
structure - for $n_l\lesssim{1}$ and $\nu=2$ one has $1-\alpha_0
\gtrsim\frac{1}{2}$ and $0<{1}-\alpha_{1\sigma}\lesssim\frac{1}{2}$
- which needs care and makes numerical calculations to higher orders
much more time consuming than NCA or SNCA, due to multiple convolutions
of these singular structures.
\begin{fmffile}{fmf_nca2}
\begin{fmfshrink}{0.8}
  \fmfcmd{%
    style_def wiggly_arrow expr p =
    cdraw (wiggly p);
    cfill (arrow p)
    enddef;
    style_def dbl_wiggly_arrow expr p =
    draw_double (wiggly p) ;
    cfill (arrow p);
    enddef;
  }

\subsection{Generalizations of SNCA: PNCA, ENCA and FNCA }

\begin{figure}[t]
  \begin{center}
    \input{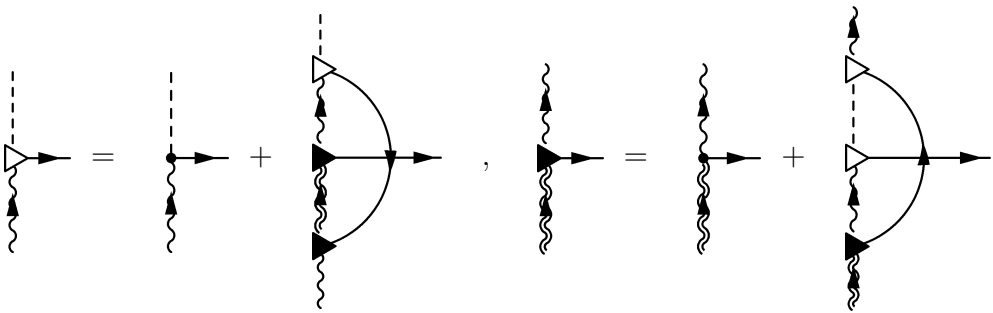}
  \end{center}
  \caption{Vertex structure of the "full NCA" (FNCA). The hierarchy of
    vertex corrections is generated by two coupled integral equations.
    The bare vertices ( first terms) are subsequently crossed by one
    more band excitation, which ends below and above in a full vertex,
    respectively. The diagrams are taken as skeletons, i.e.\ the local
    lines are dressed with the full ionic propagators.}
  \label{Figure2}
\end{figure}


The first useful generalization of NCA to the SIAM with general
values of the Coulomb repulsion $U$ was proposed and investigated in
1989~\cite{Quelle37}. It was called "full NCA" (FNCA) and is
visualized diagrammatically in figure~\ref{Figure2}. One recognizes a
particular subsystem of integral equations, which serves to generate
a class of vertex corrections (again as skeletons) extending to
infinite order. This particular choice was motivated by an attempt
to include as many as possible exchange counterparts to those
processes, which already contribute to the ionic propagators in SNCA, 
see appendix in ~\cite{Quelle37}. 
\begin{figure}[t]
  \begin{center}
    \input{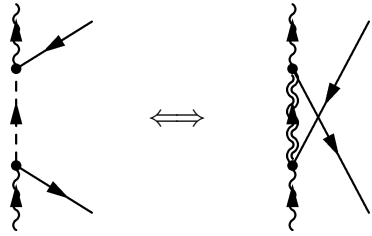}
  \end{center}
  \caption{Sequences of two elementary excitation processes, which
  together constitute the lowest order exchange coupling vertex remaining
  after a Schrieffer-Wolff transformation of SIAM to the s-d-model.}
  \label{Figure3}
\end{figure}
As Keiter repeatedly
has pointed out~\cite{Quelle38}, the balance between processes which
transform into each other by a reversal of partial time orderings,
as shown in figure~\ref{Figure3}, is necessary to obey the
Pauli-principle and to comply with universality in the Kondo limit,
where in accord with the Schrieffer-Wolff transformation from SIAM
to the s-d-exchange model~\cite{Quelle39} the characteristic energy
scale $k_B{T_K}$ is expressed via an effective exchange coupling
constant $I=\frac{V^2}{\epsilon_\ell}-\frac{V^2}{\epsilon_\ell+U}$;
figure~\ref{Figure3} just visualizes the two contributions to this $I$
~\cite{Quelle11,Quelle40}.

The system of five integral equations
according to figures~\ref{Figure1} and~\ref{Figure2} was solved for
finite $U$ in ~\cite{Quelle37}, and the results for the ionic
propagators and the corresponding excitation spectra were compared
to some simpler calculation schemes. Whereas pronounced
discrepancies to SNCA showed up, e.g.\ regarding the important
energy scales, the so called "enhanced NCA" (ENCA) already captured
important improvements.

In ENCA all vertices on the right hand side
of the two equations in figure~\ref{Figure2} are taken as bare ones.
Then, only the leading contributions to the infinite series of
vertex corrections contained in FNCA are included; among the latter
are running n-particle cascades between initial and final state
during the excitation by the external electron (iterate the vertex
in the middle of the last diagram) as well as long-time memory
effects between initial and final states through chains of internal
excitations, before or after the external excitation occurs (iterate
the respective vertices on top and at the bottom of the diagram).
Since the ENCA has proven as a good compromise between accuracy and
the calculational effort to be invested in an impurity solver for
lattice problems (see also section 4), we cite the explicit
expressions for the vertex corrections, which have to be solved
together with the system (\ref{Gleichung11}) of self-energy
equations:
\begin{eqnarray}
\Delta\Lambda^{(ENCA)}_{0,1\sigma}(z,z')&=\int{d\epsilon{D_{-\sigma}}}(\epsilon)f(\epsilon)
P_{1-\sigma}(z+\epsilon)P_2 (z+z'+\epsilon),\nonumber\\
\Delta\Lambda^{(ENCA)}_{2,1-\sigma}(z,z')&=\int{d\epsilon{D_{-\sigma}}}(\epsilon)(1-f(\epsilon))
P_{1\sigma}(z-\epsilon)P_0(z-z'-\epsilon). \label{Gleichung16}
\end{eqnarray}
Calculations of the local one-particle spectrum in
~\cite{Quelle37} were then based on the ENCA and led to an
improved many body scale and a better understanding of the many body
dynamics of SIAM, in particular at finite values of $U$.

Whereas ENCA takes into account the vertex corrections up to order
$\mathcal{O}(V^2)$, and FNCA in addition certain classes up to infinite order,
both do not include the fully crossing diagram of order $\mathcal{O}(V^4)$
shown in figure~\ref{Figure4}(a). 
\begin{figure}[t]
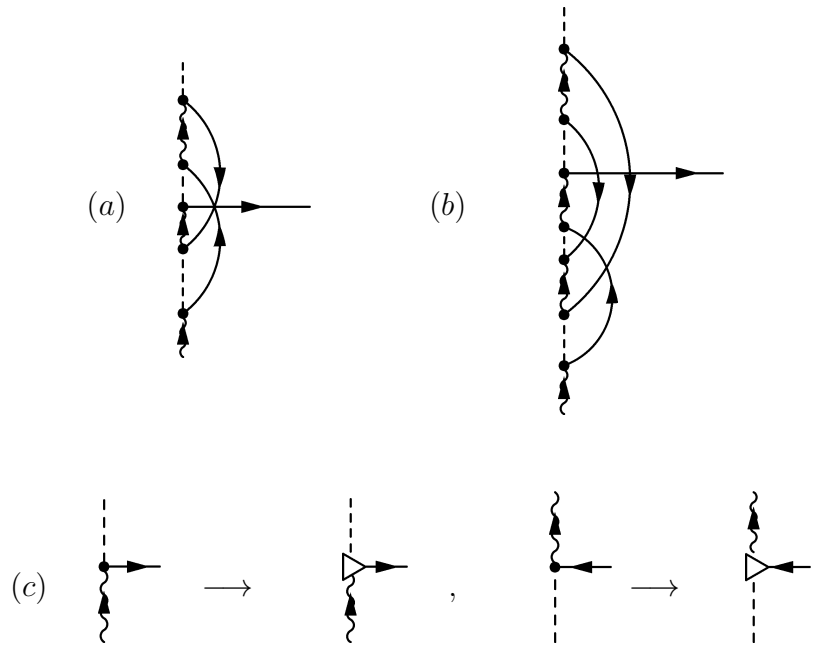

  \begin{center}
    \input{diag_enca.tex}\\[1cm]
    \input{replace_nca.tex}
  \end{center}
  \caption{Part (a) shows a fully crossing vertex correction of order $\mathcal{O}(V^4)$.
  The two processes shown in parts (a) and (b) are of the same order regarding an expansion in the
  degeneracy $\nu$ of the singly occupied local state. With the self consistent replacements shown in
  part (c) and with fully dressed local lines they constitute the "post-NCA" (PNCA), a theory for $U=\infty$,
  in which the doubly occupied local state is projected out.}
  \label{Figure4}
\end{figure}
This vertex correction is the lowest
non vanishing one in the $U=\infty$ -theory and was frequently used
to discriminate "crossing" and "non crossing" approximations.

In order to investigate the role of such fully crossing diagrams the
SIAM at infinite $U$ was investigated in 1994 with help of a
"post-NCA"(PNCA)~\cite{Quelle41}. This approximation scheme was set
up along the lines of a $\frac{1}{\nu}$-expansion and collected all
vertex corrections up to $\mathcal{O}\left(\frac{1}{\nu^2}\right)$, i.e.\ all
contributions to the ionic self energies up to this order.
Therefore, also the vertex correction shown in figure~\ref{Figure4}(b)
was taken into account, which has two more powers of $V$ compared
with figure~\ref{Figure4}(a), but due to
$\sum\limits^\nu_{\sigma'=1}\:\left(\frac{V}{\sqrt{\nu}}\right)^6=\frac{V^6}{\nu^2}$
is of the same order $\frac{1}{\nu^2}$ thanks to the closed ring
with spin-summation over $\sigma'$ between vertices 2, 3, 6, and 7.
Actually, and in close analogy to the FNCA, the bare vertices in
figure~\ref{Figure4}(a) and (b) were all replaced by full ones, as
indicated in figure~\ref{Figure4}(c), and the coupled system of vertex
corrections (now including all orders) was solved, again
self-consistently together with the system (\ref{Gleichung11}) of
ionic self energies. Convergence could be reached on not too large
time scales by use of parallel computing. Progress over the
original NCA turned out to be essential: Apart from a corrected
many-body scale $k_B{T_K}$, the local Fermi-liquid properties
improved considerably, the position of the ASR near the Fermi level
agreed much better with the one implied by Friedels sum rule, and
also the threshold exponents $\alpha_0$ and $\alpha_1$ were shifted
towards the values of~(\ref{Gleichung15}), although agreement
with these values or with a variant according to
~\cite{Quelle42} was not conclusive.

Due to the considerable
numerical effort, regarding the multiple overlapping integrations
over functions with rich structure, an extension of PNCA to finite
values of $U$ seemed not possible in 1994, since many more diagrams
involving the doubly occupied state would have to be added.

\bigskip

\subsection{Other approximation schemes in the literature}
Before the new approximations CA1 and a CA2-project (in section 5)
will be explained, we shortly comment on two approximation schemes,
which have been proposed and investigated over the last ten years.
In the so called "symmetrized finite-$U$ NCA" (SUNCA) special
emphasis is laid on the chains of scattering events~\cite{Quelle43}
mentioned above in connection with the FNCA~\cite{Quelle37}. This
scheme is conserving ($\Phi$-derivable in the sense of Kadanoff and
Baym) like all other approximations mentioned in this section;
moreover it can be characterized as involving just a subclass of the
FNCA-diagrams. Although the relevant papers are written with help of
the slave-boson formalism, the formulation is fully equivalent to
direct perturbation theory as pointed out above. The evaluation of
the local one particle spectrum is based on a full infinite subclass
of vertex corrections and thus goes beyond the ENCA-calculations.
These vertex corrections with long scattering chains are easily
visualized with help of the FNCA-diagrams of figure~\ref{Figure2}:
Iterate the vertex equations with respect to the upmost vertex only.
The results underline the progress reached with ENCA and
FNCA~\cite{Quelle37}.

\begin{figure}[t]
    \input{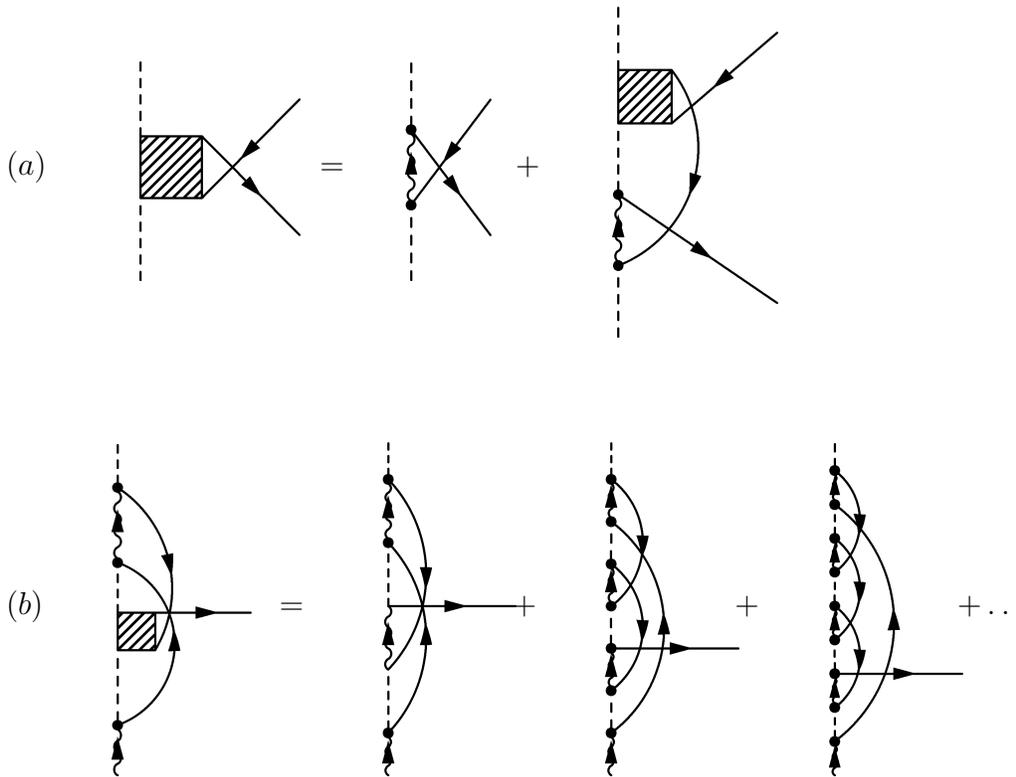}
  \caption{Vertex corrections summed in the "conserving T-matrix approximation"
  (CTMA) for the SIAM at $U=\infty$ as viewed from direct perturbation theory.
  Part (a) T-matrix which, when substituted into the diagram of
  figure~\ref{Figure4}(a), generates the sequence of vertex corrections shown in
  part (b).}
  \label{Figure5}
\end{figure}
Whereas SUNCA is applicable to the finite-$U$ case and can be placed
into a scheme of repeated vertex corrections with single line
crossings (a more general version being FNCA), the "conserving
$T$-matrix approximation" (CTMA) ~\cite{Quelle44} again is restricted
to infinite $U$ and stresses the importance of chains of scattering
events for band electrons off the local shell over the whole
duration of the external excitation process. These are argued to
contain those significant contributions, which are known to lead to
the correct singular threshold behaviour of $X$-ray absorption
spectra as predicted by Mahan~\cite{Quelle45} and calculated by
Nozieres et al.~\cite{Quelle16}.

Correspondingly, essentially exact
threshold exponents are expected from the CTMA. This approximation
can be characterized with reference to the fully crossing diagram of
figure~\ref{Figure4}(a): The middle part between vertices 2 and 3
becomes the lowest contribution to a $T$-matrix, which is fully
determined by the implicit equation shown in figure~\ref{Figure5}(a).
It generates the sequence of vertex corrections with scattering
chains shown in figure~\ref{Figure5}(b). Observe that only the first
of these is contained in PNCA.

In spite of a superficial resemblance
already the second contribution is different from
figure~\ref{Figure4}(b), which is more easily recognized by counting
the number of independent spin-summations. Indeed,
CTMA-results~\cite{Quelle46} point to considerably improved values
of the threshold exponents; nevertheless, the description of the
local Fermi-liquid formation, similar to PNCA, is still not fully
satisfactory. Both of these approximations involve time-consuming
numerical calculations; up to now a generalization to the even more
demanding case of finite $U$-values has not been reported.

Other approximation schemes involving additional simplifying assumptions 
for the ionic propagators and vertex functions, be it either 
in  a non-conserving~\cite{Quelle46a} or conserving fashion~\cite{Quelle46b},
will not be considered here. 
Although they may be useful  with respect to computational effort, they
have only been justified for the case of large orbital degeneracy.

\subsection{CA1 approximation}

\begin{figure}[t]
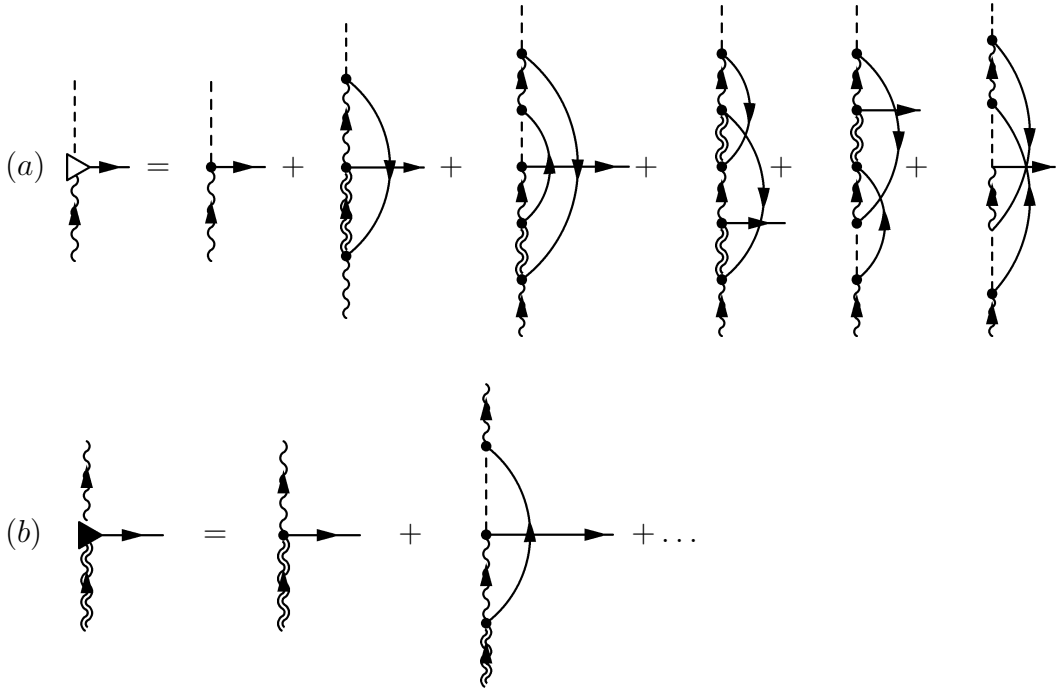

 $(a)$\input{ca1a.tex}\\[6mm]
 $(b)$\input{ca1b.tex}\\[3mm]
  \caption{CA1 collects all vertex corrections for general (finite
    and infinite)
    values of $U$ up to order $\mathcal{O}(V^4)$; in part (a) these are shown explicitly for
    one of the two vertices. The analogous construction for the other vertex is
    indicated in part (b) by dots. Again local lines are dressed, i.e.\ the diagrams
    are used as skeletons; the vertex points on the right hand side, however, are
    bare ones.}
  \label{Figure6}
\end{figure}
\bigskip

CA1 is designed to describe SIAM in the full range of values for the
local Coulomb repulsion $U$ with good accuracy and likewise for
dynamical properties at general excitation energies $\omega$. Being
a straightforward collection of all vertex corrections up to order
$\mathcal{O}(V^4)$ (as skeletons) it is conserving and contains the leading
contributions from all of the approximations sketched above. More
precisely, it can be defined by the set of vertex corrections shown
in figure~\ref{Figure6}(a), plus the corresponding ones for the other
vertex, indicated by points in figure~\ref{Figure6}(b). In successive
order the diagrams may be characterized as follows: The first terms on
the r.h.s. are the bare
vertices and define SNCA. ENCA additionally contains the following
vertex correction with a single crossing electron or hole line,
respectively. The next three vertex corrections can be viewed as
originating from the ENCA-diagram by dressing each of the vertices
with a single crossing line successively, i.e.\ they represent the
first iteration in the FNCA-scheme. The last diagram is the
fully-crossing contribution not contained in the FNCA; it is the
leading vertex correction in both approximations for infinite $U$,
PNCA, and CTMA.

Whereas CA1 is explored in the following together
with the other approximations mentioned, a CA2-project will be
designed to add more vertex iterations like those included in FNCA
and longer scattering chains like those of the CTMA, which can be
incorporated via a T-matrix formalism.
\section{Results from CA1 and comparison with other impurity solvers}
\subsection{General remarks on the quality of impurity solvers}
The following criteria have frequently been applied to judge the
quality of impurity solvers in connection with the Anderson impurity
model: (1) Ionic propagators have to obey the correct threshold
behaviour, in accord with the relevant work on orthogonality
catastrophy and excitonic correlations in the case with
spin-degeneracy~\cite{Quelle17,Quelle45,Quelle47,Quelle48}. (2) The
infrared divergencies of the perturbation series produce a
characteristic low energy scale, usually referred to as the
Kondo-temperature $T_K$, which should faithfully be reproduced by
the approximation. (3) The many-body resonance
(Abrikosov-Suhl-resonance, ASR) forming at temperatures of order of
the Kondo-scale $T_K$ and lower is pinned at a position near the
Fermi level, which is determined by Friedel큦 sum rule. (4) The
shape of the ASR has to comply with a form of the selfenergy, which
guarantees local Fermi-liquid properties.

In the following, the four
semianalytical impurity solvers introduced hitherto for the SIAM
with general, in particular finite values of $U$, i.e.\ SNCA, ENCA,
FNCA, and SUNCA will be compared to the new CA1. Special emphasis
will be laid on the above four criteria. As a reference, also
NRG-calculations for the SIAM are presented, which in the low-energy
regime should provide a reliable bias. They have been produced with
help of the very effective numerical procedures presented
in~\cite{Quelle49}. All other calculations have been performed with
a software package written for the solution of a number of impurity-
and lattice-models, in which various impurity solvers can be
combined with different methods for the lattice aspects. Since it is
based on adaptive strategies for an all-purpose use, no particular
provisions have been taken to optimize numerical strategies for the
SIAM in the deep Kondo limit. Nevertheless, the program package
seems to work very reliably, although numerical convergence problems
and approximation errors become visible for certain extreme choices
of model parameters. In particular, the folding of several nearly
singular factors in an integrand like that of~(\ref{Gleichung12})
 at very low temperatures, needs a thorough
analytical preparation and consumes much numerical effort, and
likewise the iteration of vertex parts depending on two
energy-variables for SUNCA, FNCA, and CA1.

\subsection{Ionic threshold behavior}
The shortcomings of SNCA have already been mentioned in the
foregoing section. It proves worthwhile, however, to check how the
known values of threshold exponents for this approximation are
recovered in the calculations; this gives valuable hints as to how
results for the other approximations should be interpreted and
generally, how reliably the algorithms work. It is interesting to
note here, that the information given about the NCA-threshold
exponents (i.e.\ the case $U=\infty$) in connection with
~(\ref{Gleichung15}) is not complete when regarding SNCA at
finite $U$. The treatment of ~\cite{Quelle15} based on an ansatz
for ionic selfenergies is easily generalized:
\begin{eqnarray}
\fl
\Sigma_M(\omega+i\delta)\approx{E_g}-E_M-\frac{i}{A_m}(\omega-E_g)^{1-\alpha_M}\rightarrow{P_{n_0M}}
(\omega+i\delta)\approx-iA_M(\omega-E_g)^{\alpha_M-1}
\label{Gleichung17}
\end{eqnarray}
It furnishes the same asymptotic region near $\omega=E_g$ as in the
case $U=\infty$, characterized by
$\alpha_0=\frac{1}{3}\:,\:\alpha_{1\sigma}=\frac{2}{3}$ for (spin-)
degeneracy two, for the whole unsymmetric regime
$(\epsilon_{\ell}\equiv{E}_{1\sigma}-E_0<-\Delta\:\:\mathrm{and})\:\:2\epsilon_{\ell}+U>0$.
Asymptotically, the propagator $P_2$ does not contribute here. For
$U$ approaching the value $-2\epsilon_{\ell}$ from above, however,
$P_2$ becomes equal to $P_0$, which leads to
\begin{eqnarray}
\alpha_1=\frac{\nu}{2+\nu}\:,\quad\alpha_0=\alpha_2=\frac{2}{2+\nu}\quad(2\epsilon_{\ell}+U=0),
\label{Gleichung18}
\end{eqnarray}
$\nu$ being the degeneracy of ionic state $|1\sigma\rangle$. For the
model with spin-degeneracy only this means
$\alpha_0=\alpha_1=\alpha_2=\frac{1}{2}$. These values coincide with
the presumably exact ones taken from ~(\ref{Gleichung15}).
Naturally this does not imply that SNCA becomes correct for the
symmetric SIAM. As we will see below, e.g.\ the shape of the ASR
still reveals serious shortcomings. The transition from $U=\infty$
to $U=-2\epsilon_{\ell}$ happens in a gradual way: The former
asymptotic regime around $\omega=E_g$ shrinks to zero, and a
different regime takes over, which originally was situated at higher
values of $|\omega-E_g|$ and developed the exponents of
~(\ref{Gleichung18}). 
\begin{figure}[t]
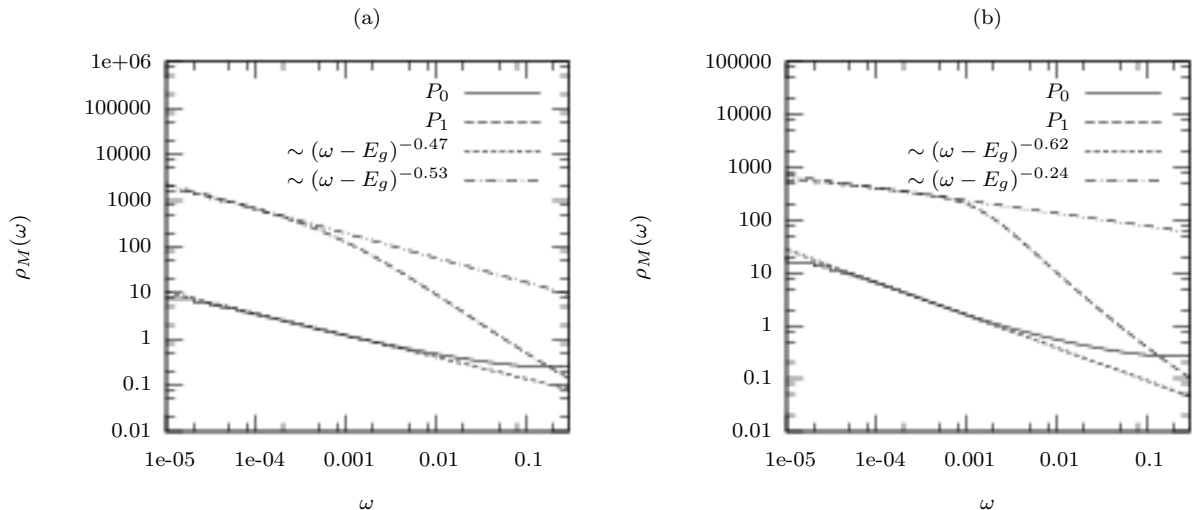

  \begin{center}
    {  \scriptsize
      \hspace*{-1.5cm}\input{Fig_7a.tex}\input{Fig_7b.tex}
    }
\end{center}
  \caption{Double-logarithmic plot of ionic spectra, centered at the
    threshold, for a SIAM
    in SNCA (a) in the symmetric case with $\epsilon_\ell=-1.0,\:U=2.0,\:\beta=\infty,$ (b)
    in an asymmetric case with $\epsilon_\ell=-1.0,\:U=3.0,\:\beta=\infty,$ both for an Anderson width
    $\Delta_A\equiv\pi{V^2}\varrho^{(0)}_{c\sigma}(0)=0.3$ and a $3d-sc$ band density of states
    $\varrho^{(0)}_{c\sigma}(\omega),$ centered at $\omega=\mu=0;$ the bandwidth is 6 units.
    The threshold exponents can be read off as one plus the slope of the asymptotic tangents drawn
    in the Figures.}
  \label{Figure7}
\end{figure}

In figure~\ref{Figure7} and~\ref{Figure8} we
show the results of a rather precise SNCA-calculation of a symmetric
SIAM in the deep Kondo regime, parts (a), and an asymmetric one,
parts (b). The impurity states locally hybridize with a
tight-binding simple cubic conduction band in three dimensions of
width 6, Fermi level and band center lie at energy $\omega=0$, and
van-Hove-singularities at $\omega=\pm1$. Shown in figure~\ref{Figure7}
are the spectra of the relevant ionic propagators $P_0$ and
$P_{1\sigma}$ for $T=0$ on a doubly logarithmic scale with origin at
the corresponding threshold energies $E_g$. Numerical resolution is
somewhat below $10^{-4}$, acting as an effective temperature cutoff.
The asymptotic regime is entered only one order of magnitude higher,
at about $\omega-E_g\approx10^{-3}$. From the slope of the tangents
drawn one reads off the exponents
$\alpha_0\approx{0.53},\:\alpha_1\approx{0.47}$ in the symmetric
case, and $\alpha_0={0.38},\:\alpha_1={0.76}$ in the asymmetric one.
Even if by a proper extrapolation, using several low values of the
temperature, these numbers can be brought closer to the exactly
known ones, i.e.\ $(0.5,\:0.5)$ and $(0.33,\:0.67)$, an uncertainty of
about 10 percent remains. This is enough, however, to identify and
discriminate the two different situations. With lower numerical
accuracy or at higher temperatures $T>10^{-4}\approx{T_K/200}$ the
asymptotic regime will be hard to attain.

\begin{figure}[t]
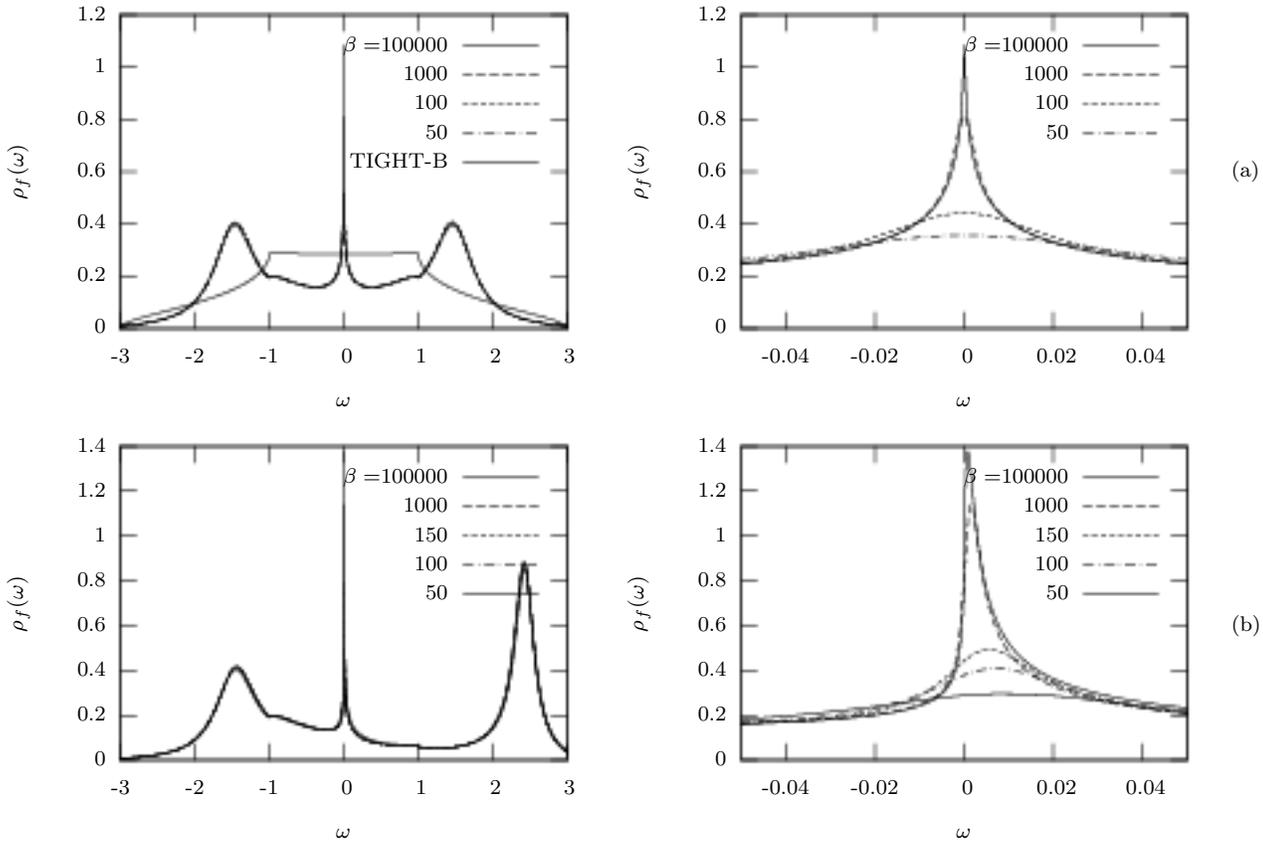

  { \scriptsize
    \hspace*{-1.5cm}\input{Fig_8a.tex}\input{Fig_8ah.tex}
    \hspace*{-1.5cm}\input{Fig_8b.tex}\input{Fig_8bh.tex}
  }
  \caption{One-particle excitation spectrum of SIAM in SNCA, parameter values as in figure~\ref{Figure7},
    $\beta$-values as specified.}
  \label{Figure8}
\end{figure}
Figure~\ref{Figure8}(a)
demonstrates that in spite of accurate threshold exponents the shape
of the ASR at $\omega=0$ in the 1-particle-spectrum for low $T$
comes out as a quite unphysical spike. Part (b) of this figure shows
how the ASR is deformed, when $U$ is raised; the position of the ASR
is clearly temperature dependent, and its flank at $\omega=0$
develops a pathological steepness. This reminds of the pathological
structure found at infinite $U$ with a flat conduction band density
of states~\cite{Quelle15}. Since for our calculations a three dimensional
simple-cubic tight-binding bandstructure was used, van-Hove singularities 
(i.e.\ kinks at $\omega=\pm 1$) leave their traces in the spectrum;
their visibility also constitutes a test for the 
numerical procedures used.
\begin{figure}[h!]
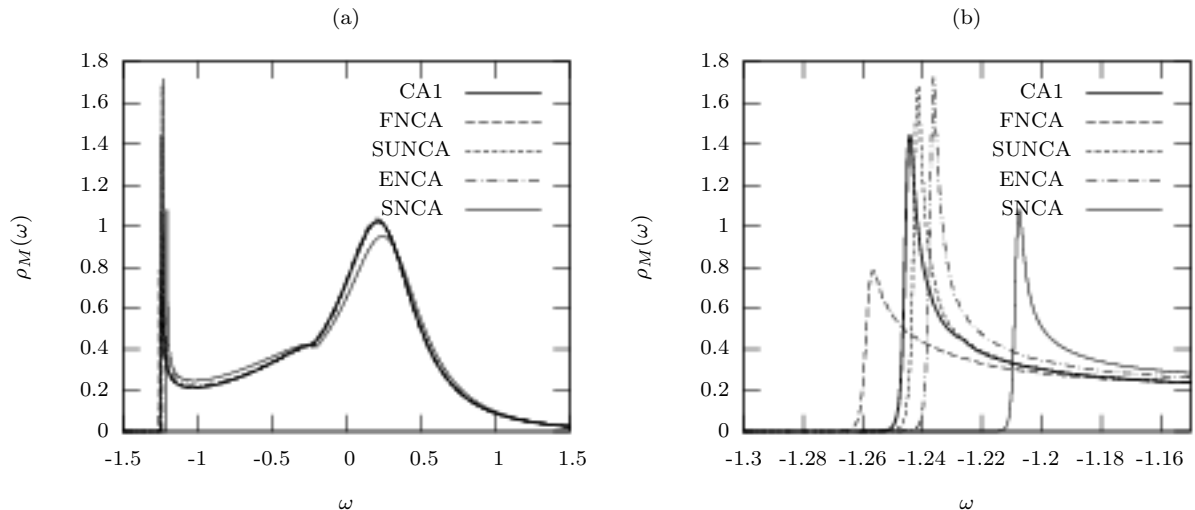

  \begin{center}
    { \scriptsize
    \hspace*{-1.5cm}\input{Fig_9a.tex}\input{Fig_9b.tex}
  }
\end{center}
  \caption{Spectral density of the empty ionic state $M=0$ for a SIAM,
    calculated with the five semianalytical approximations discussed in
    the text, parameter values as in figure~\ref{Figure7}, $\beta=1000$. The shifted
    thresholds allow for fits of different quality to the Kondo
    temperature $T_K$.}
  \label{Figure9}
\end{figure}
\begin{figure}[h!]
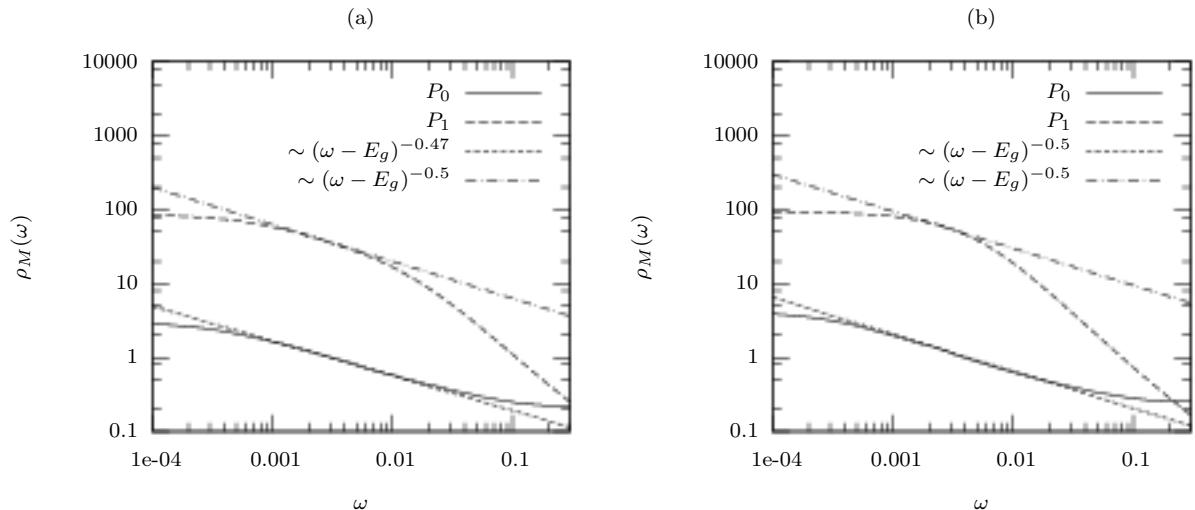

  \begin{center}
    {  \scriptsize    
      \hspace*{-1.5cm}\input{Fig_10a.tex}\input{Fig_10b.tex}
    }
  \end{center}
  \caption{Double-logarithmic plot of ionic spectra, centered at the
    threshold, for a SIAM in ENCA, parameter values as in figure~\ref{Figure7}.}
  \label{Figure10}
\end{figure}
\begin{figure}[h!]
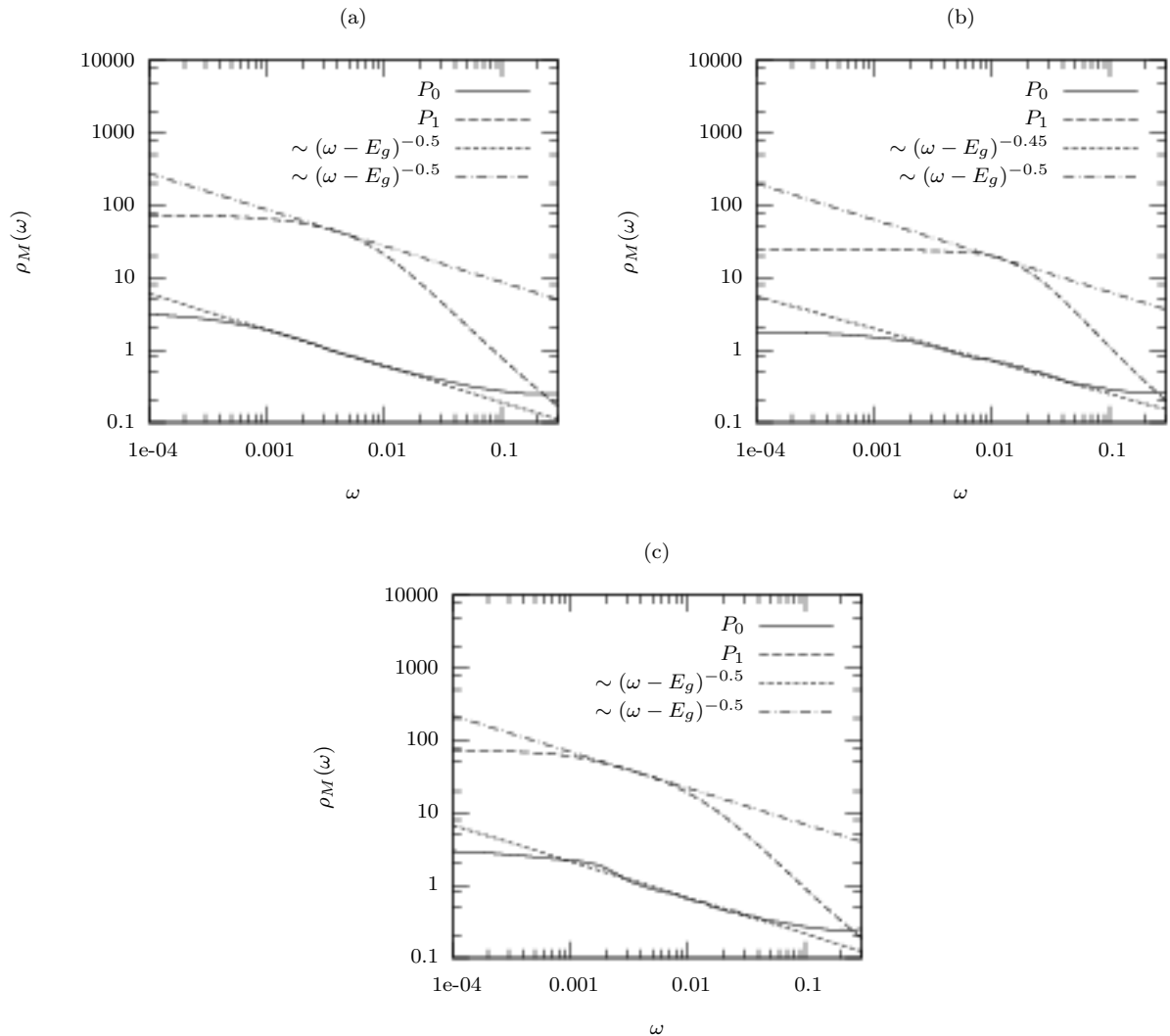

  \begin{center}
    {  \scriptsize
      \hspace*{-1.5cm}\input{Fig_11a.tex}\input{Fig_11b.tex}
      \hspace*{-1.5cm}\input{Fig_11c.tex}
    }
  \end{center}
  \caption{Double-logarithmic plot of ionic spectra, centered at the
    threshold, for a SIAM in SUNCA (part (a)), FNCA (part (b)) and CA1
    (part (c)), all for the asymmetric case with parameter values as in
    figure~\ref{Figure7}.}
  \label{Figure11}
\end{figure}
\begin{figure}[h!]
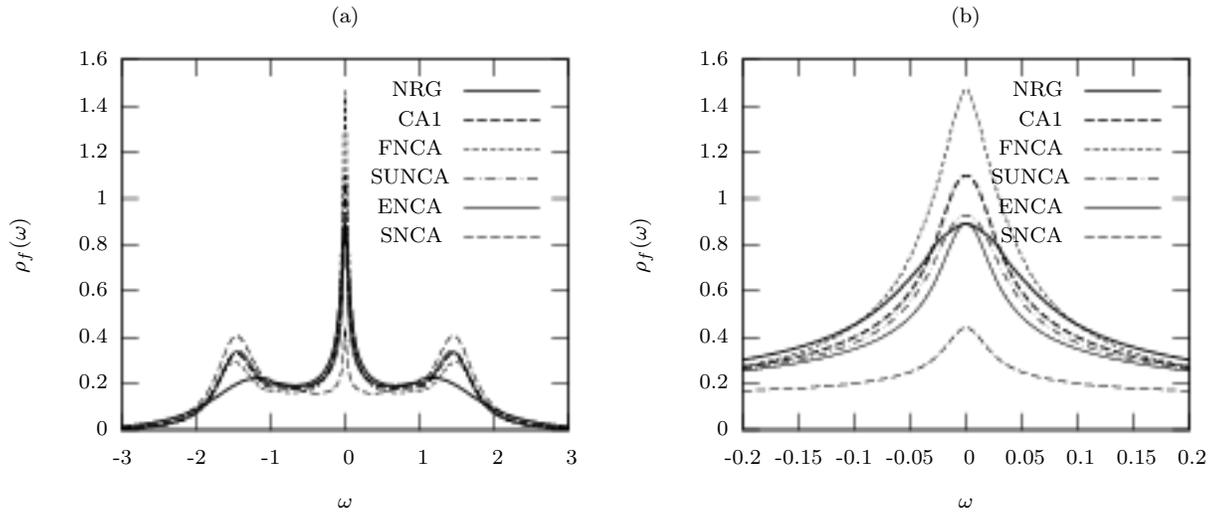

  \begin{center}
    {  \scriptsize
    \hspace*{-1.5cm}\input{Fig_12a.tex}\input{Fig_12b.tex}
  }
  \end{center}
  \caption{One-particle excitation spectrum of a symmetric SIAM at
    $\beta=100$, other parameters as in figure~\ref{Figure7}(a), in the five
    semianalytical approximations discussed in the text and,
    additionally, calculated with the numerical renormalization group
    (NRG). Part (a) reveals shortcomings of the NRG-method at large
    excitation energies, whereas in the low-energy region of part (b)
    the NRG-curve can be used as a reference for the other
    approximations. } \label{Figure12}
\end{figure}
\begin{figure}[h!]
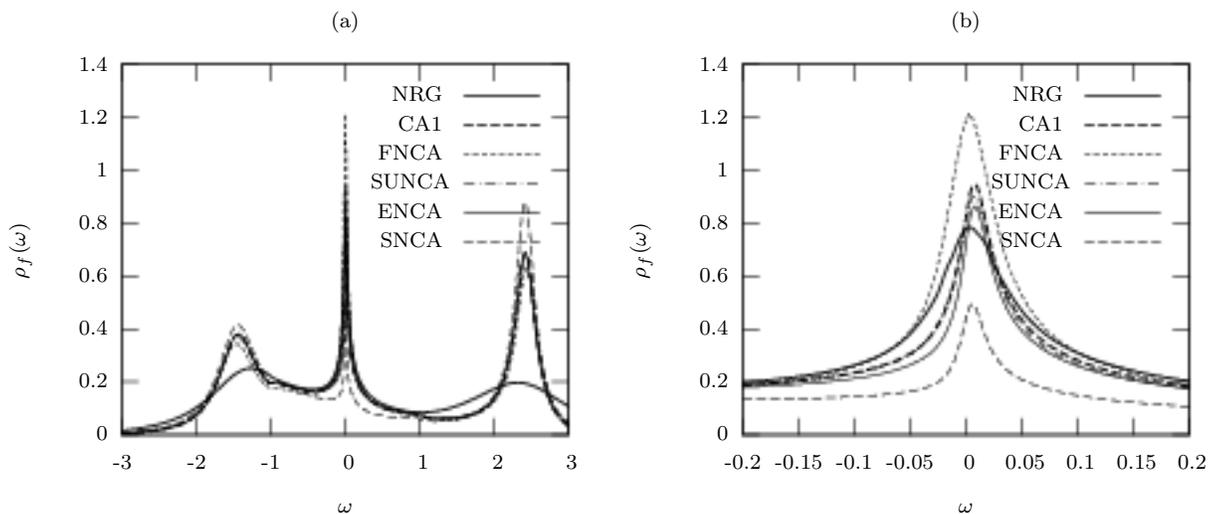

  \begin{center}
       {  \scriptsize
         \hspace*{-1.5cm}\input{Fig_13a.tex}\input{Fig_13b.tex}
       }
  \end{center}
  \caption{Analogue of figure \ref{Figure12} for the asymmetric case with
    $\beta=150$, other parameters as in figure~\ref{Figure7}b. }
  \label{Figure13}
\end{figure}
\begin{figure}[h!]
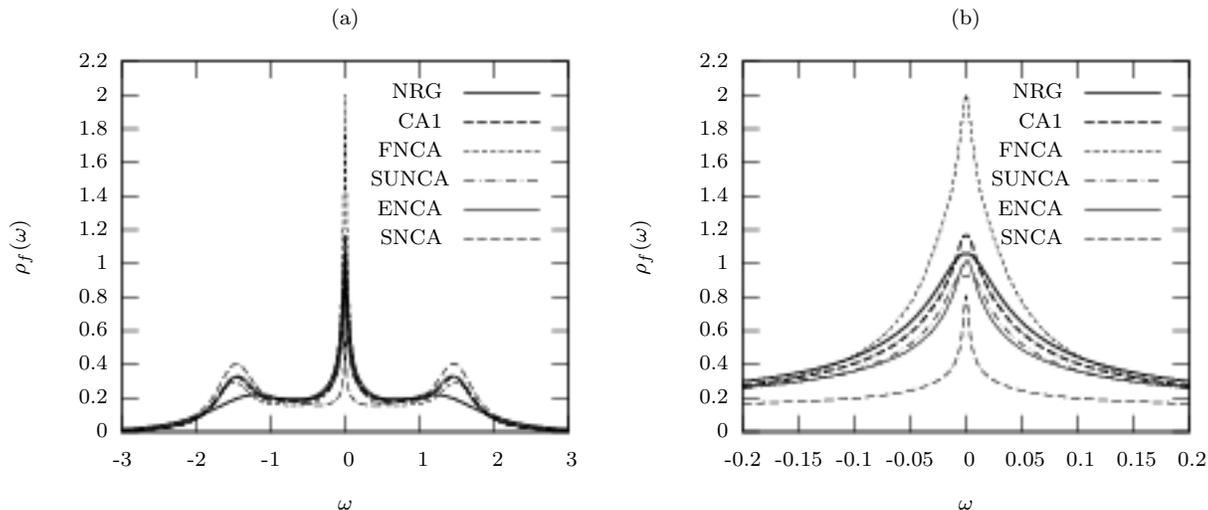

  \begin{center}
    {  \scriptsize
    \hspace*{-1.5cm}\input{Fig_14a.tex}\input{Fig_14b.tex}
  }
  \end{center}
  \caption{Like figure~\ref{Figure12}, but with $\beta=1000$, i.e.\ at $T\ll T_K$.  }
  \label{Figure14}
\end{figure}
%
%
%

A good qualitative insight into the relation between the five
semianalytical impurity solvers under consideration can be obtained
from figure~\ref{Figure9}, which shows the spectrum of $P_0=P_2$ for the symmetric
SIAM discussed before at temperature $T=10^{-3}$. Whereas part (a)
gives an overall view with the threshold to the left and a broad
one-particle resonance to the right, corresponding to a distribution
of contributing frequencies around
$\omega=-E_{1\sigma}-\Delta{E_{1\sigma}}$ ($E_0$ is set to
zero in all calculations), part (b) with a much finer
energy-resolution points to the discrepancies between the different
approximations visible in the low-energy regime. One recognizes
threshold-peaks at different values of $E_g$, in increasing order
for FNCA, CA1, SUNCA, ENCA, and SNCA. Differences in $E_g$ directly
reflect the ability of the approximations to reproduce the
Kondo-scale, which in this regime can be expressed
as~\cite{Quelle50}:
\begin{eqnarray}
[k_B]T_K=a\sqrt{\mathrm{I}}\:\textrm{exp}\left[-\frac{\pi}{\mathrm{I}}\right],\quad\mathrm{I}=-
\frac{2U\Delta_A}{\epsilon_\ell(\epsilon_\ell+U)}\quad.
\label{Gleichung19}
\end{eqnarray}
Choosing $a=\frac{U}{2\pi}$ for $U\leq$
bandwidth $W$~\cite{Quelle37} we obtain $T/T_K=0.04$, i.e.\ the
spectra represent the temperature range well below $T_K$, even with
slightly different choices of the coefficient $a$. If one accepts
for the moment, that the FNCA with lowest $E_g$ furnishes the
closest approximation to the real $T_K$, as e.g.\ is implied by the
NRG-calculation (see below) then the following conclusion can be
drawn: ENCA, SUNCA, and CA1 all improve considerably on the SNCA.
The leading vertex correction already included in the ENCA
contributes the essential part to this effect, whereas the
additional terms further taken into account in SUNCA and CA1,
respectively, have a relatively smaller impact. This agrees with the
original investigation of ENCA and FNCA~\cite{Quelle37}, where it
was shown, that the ENCA already captures the right exponential
behaviour of $T_K$ for the SIAM at finite $U$, whereas the inclusion
of further vertex corrections then only improves on the prefactor in
this scale.

As will be shown below, the good estimate of $T_K$
furnished by the FNCA does not imply that FNCA behaves well in all
other respects, e.g.\ concerning the four points mentioned in the
beginning. What is apparent, however, is the pronounced and
qualitatively similar threshold behaviour visible in all of the five
approximation schemes applied to the ionic spectra.

Since ENCA, SUNCA, and FNCA all reduce to the NCA in the limit
$0>\epsilon_\ell$ fixed, $U\rightarrow\infty$, it is to be expected
that in the asymmetric case, $2\epsilon_\ell+U>0$ an ultimate
asymptotic regime very near to $\omega=E_g$ exists, where the ionic
spectra are ruled by the NCA-threshold exponents. This must not
necessarily be true for the CA1 with its fully crossing vertex
correction, which does not vanish in this limit. It can nevertheless
be anticipated, that as a precursor a regime with "better" threshold
exponents at somewhat higher values of $|\omega-E_g|$ occurs also
for ENCA, SUNCA, and FNCA. In figure~\ref{Figure10}(a) a corresponding evaluation
of ENCA is shown for a temperature which again is far below $T_K$
and also below the numerical resolution of about $10^{-3}$ or
somewhat less. The exponents for the symmetric case in figure~\ref{Figure10}(a),
as read off in the range $10^{-3}\leq\omega\leq{10^{-2}}$ are
clearly near the value $0.5$, whereas in the asymmetric case of 
figure~\ref{Figure10}(b) only $\alpha_0\approx{0.5}$ is really conclusive; a value
$\alpha_1\approx{0.5}$ can be justified only if a tangent is drawn
in the reduced range between $\omega\approx{10^{-3}}$ and
$\omega\approx0.5\cdot10^{-2}$ before the steeper decrease sets in.
Comparing figures~\ref{Figure7}(b) and ~\ref{Figure10}(b) this seems to be a
reasonable procedure. It must be remembered here that the better
approximation cannot be evaluated with the same numerical accuracy,
at least not using the program package in its present form. 
Figures~\ref{Figure11}
(a), (b), and (c) show the threshold behaviour for the asymmetric
model (again with $\epsilon_1=-1.0,\:U=3.0$ and $T\ll{T_K}$)
obtained with SUNCA, FNCA, and CA1. Similar conclusions as for the
ENCA can be drawn here: In all cases the threshold exponents in the
accessible asymptotic regime come out near the value $0.5$. At least
for the CA1 this gives reason to hope for an essential improvement
of the true asymptotics of ionic spectra over the SNCA.

\subsection{One-particle excitation spectra}

For a comparison of the one-particle excitation spectra obtained
with the five approximation schemes also a calculation with the NRG
is taken into account. This should give an impression of the exact
result, at least in a low-energy regime near $\omega=\mu=0$.

In figure~\ref{Figure12} we present the results for the symmetric SIAM
and in figure~\ref{Figure13} for the asymmetric one, at
$\beta=1/k_BT=100$ and $\beta=150$ respectively, implying $T$ being
roughly half $T_K$, other parameters as before. First looking at
parts (a) of these figures, the following fact seems remarkable:
Peaks in the NRG-spectra are considerably broader compared with the
other cases and the features due to the van-Hove singularities in
the band-DOS are smeared out much more. At least part of this, in
particular at higher excitation energies, should be due to numerical
procedures used in the NRG-calculation: It is based on a discrete
set of energy-eigenvalues, which is considerably spaced near the
bandedges and becomes logarithmically denser for smaller excitation
energy $\omega$; interpolation then has a smoothening effect.
Whereas near the original resonances at $\omega\leq\epsilon_\ell$
and $\omega\geq\epsilon_\ell+U$ the five semianalytical
approximations are certainly closer to the truth than the
NRG-curves, the situation is not completely clear in the energy
region around the ASR, although here the NRG is most trustworthy.

The peak value close to $\omega=0$ at $T=0$ is given via Friedel큦
sum rule to be $\varrho_{f\sigma}(\omega)=1/\pi\Delta_A\approx1.06$;
this is faithfully reproduced by the NRG, whereas the width of the
ASR might already be somewhat exaggerated by the NRG. In effect,
however, we take the NRG-ASR as our measure of quality for the other
approximations with respect to the low-energy regime. Regarding the
full range of excitation energies, on the other hand, any of the
other approximations (except SNCA) might be more appropriate,
depending on the purpose of the calculation. With the halfwidth of
the ASR taken as measure for the many-body scale $T_K$, certainly
the FNCA, with its vertex-corrections systematically iterated
through all orders, compares most favourably with the NRG. On the
other hand, the FNCA-spectrum clearly exaggerates the height and
thus the total weight of the ASR: The limit of $1/\pi\Delta_A$ for
the peak-height becomes violated even stronger for lower
temperatures. Insofar, CA1 seems to represent a good compromise and
even ENCA does not work too bad.

The lesson to be learned from these results is that improved
semianalytical impurity solvers of this type should incorporate
skeleton diagrams of two types in a well-balanced way: Classes of
iterated vertex corrections have to be accompanied by chains of
iterated particle-scattering events, being related to each other as
exchange-partners~\cite{Quelle37,Quelle44,Quelle46}. In light of the
discussion in section 2, CA1 serves as a further step in this
direction. Regarding the regime at high excitation energies
$\omega\approx\epsilon_\ell+U$ the comparison of
figures~\ref{Figure12}(a) and \ref{Figure13}(a) reveals a trend, which
for even larger values of $U$ becomes more and more pronounced and
which apparently is not well captured by the NRG, at least with its
present numerical performance: The resonance due to double occupancy
of the local shell becomes sharper with increasing $U$, in
particular when $\epsilon_\ell+U$ reaches the order of the upper
band edge $\omega\approx\frac{1}{2}W=3$. Beyond this value, the peak
keeps its weight but rapidly looses its width and finally vanishes
as a single spectral line out of the accessible region. This is
faithfully reproduced by any of the five semianalytical impurity
solvers under consideration.

Figure~\ref{Figure14} gives an
impression about qualities and failures of the five approximations
as applied to the full calculation of one-particle spectra at very
low temperatures: $\epsilon_\ell=-1.0,\quad{U}=2.0$ and
$\beta=1000$, i.e.\ $T/T_K=0.04$ have been chosen here. Whereas the
ionic spectra in all five cases come out rather reliably with the
procedures used in our program package, the subsequent folding of
ionic propagators and defect propagators (see e.g.\
~\cite{Quelle14}) can produce spurious results near $\omega=0$.
With very sharp thresholds in all quantities at low $T$ slight
displacements of the maxima (as a consequence of numerical
procedures and rounding errors) can have a large effect on the
integrals containing several of these quantities. Although
figure~\ref{Figure14}, too, supports the conclusions drawn before, the
SUNCA-curve and to a somewhat lesser degree the FNCA-curve, show a
spurious double-peak structure near $\omega=0$, supposedly due to
such threshold-shifts. In addition, the FNCA-curve should not be
taken too seriously very close to $\omega=0$, although its shape is
in accord with the numerically more precise SNCA-calculation in
figure~\ref{Figure8}(a). FNCA overestimates the ASR-peak height
strongly, whereas CA1, in spite of a too high peak value, rather
favourably compares with the NRG-curve.

\begin{figure}[t!]
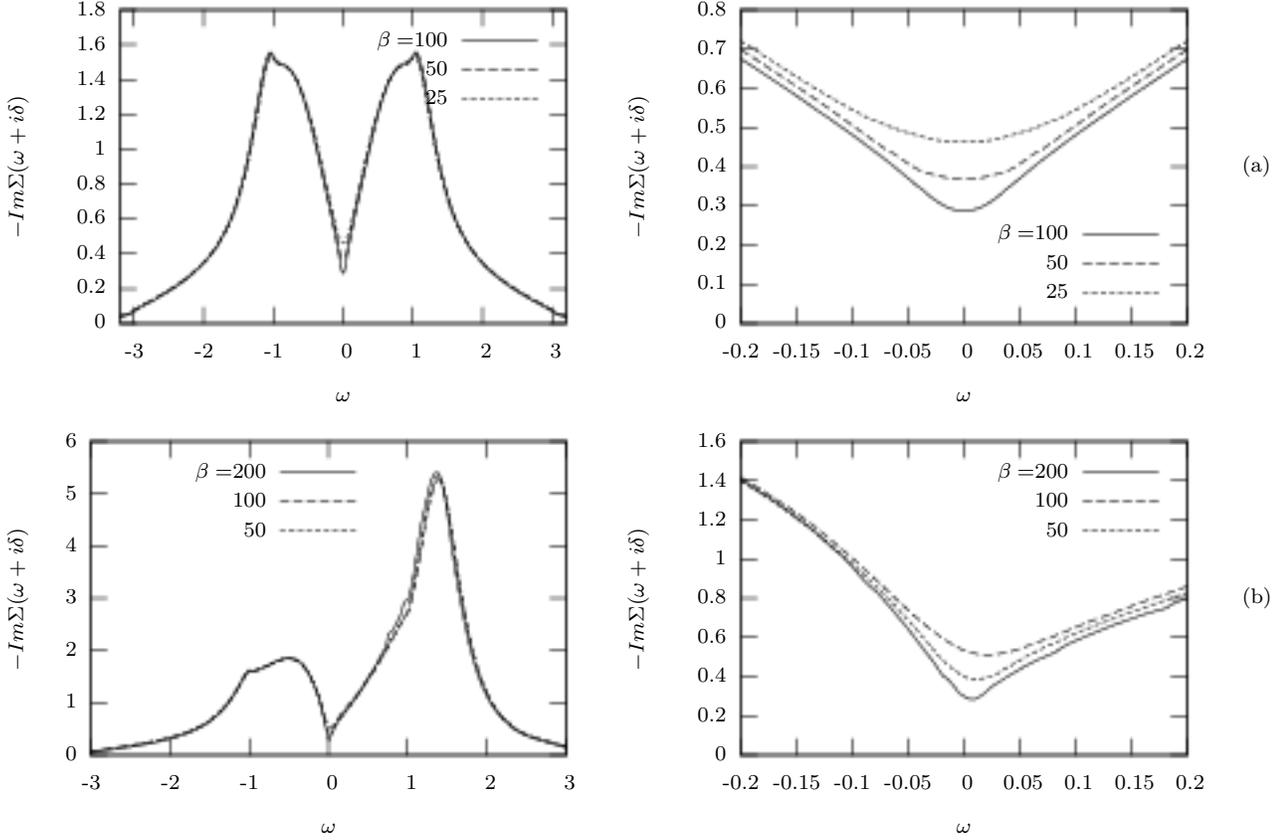

  \begin{center}
    {  \scriptsize
      \hspace*{-1.5cm}\input{Fig_15a.tex}\input{Fig_15ah.tex}
      \hspace*{-1.5cm}\input{Fig_15b.tex}\input{Fig_15bh.tex}
    }
\end{center}
  \caption{Imaginary part of the selfenergy of local electrons
    (absolute value) for a SIAM, calculated within CA1, as test for
    local Fermi-liquid properties. Temperatures as shown, other
    parameters as in figure~\ref{Figure7}.} \label{Figure15}
\end{figure}

\subsection{Fermi-liquid properties}
As a final test for our new CA1 the local Fermi-liquid properties
are inspected in figure~\ref{Figure15}. For this purpose the imaginary part of the
selfenergy
\begin{eqnarray}
-\textrm{Im}\:\Sigma_{f\sigma}(\omega+i\delta)=\frac{\pi\varrho_{f\sigma}(\omega)}
{(ReG_{f\sigma}(\omega))^2+(\pi\varrho_{f\sigma}(\omega))^2}
\label{Gleichung20}
\end{eqnarray}
is shown for a few temperatures near and well below $T_K$. The
formation of a minimum at $\omega=0$ obviously takes place. In the
asymmetric case of figure~\ref{Figure15}(b) a displacement of this
minimum away from the Fermi level with growing temperature is
recognized, similar but somewhat weaker than has been reported
before for (S)NCA and ENCA~\cite{Quelle37}, as well as for the PNCA,
the latter being a $U=\infty$-theory with crossing contributions to
vertex corrections in high orders~\cite{Quelle41}. The CA1-result
for the value of -Im$\:\Sigma_{f\sigma}(\omega+i\delta)$ at
its minimum falls short of the exact limiting value
$(\pi\varrho_{f\sigma}(\omega))^{-1}=\Delta_A=0.3$ for the lowest
temperatures. This is in accord with the too high ASR-value seen
e.g.\ in figure~\ref{Figure14}. In a quantitative sense, however, this
result improves on ENCA and, much more, on (S)NCA. Obviously, the
minimum can well be fitted by a parabola as long as the temperature
does not become so low, that numerical deficiencies near $\omega=0$
become predominant. Also its position and height, as well as the
quadratic coefficient may be compared to exact results for
Fermi liquids~\cite{Quelle11,Quelle56}, e.g.\ as a guideline for
corrective measures when using these impurity solvers in
lattice-calculations. These conclusions are similar to those for the
PNCA published before~\cite{Quelle41}. Furthermore, it is a
remarkable fact how dramatic the scattering rate raises for
increasing excitation energies. In the range of the ionic resonances
$\omega=\epsilon_\ell$ and $\omega=\epsilon_\ell+U$ it becomes high
enough to completely prevent locally a band picture even for the
c-electrons. This will become even more evident in the lattice
calculations of the next section, where the effect 
occurs on each lattice site and thus 
affects the whole Bloch-states.
\section{Impurity solvers and lattice theories}

\subsection{Cumulant perturbation theory of strongly correlated lattice models}

The importance of impurity solvers for approximate solutions of
lattice problems came to light in theories like ATA~\cite{Quelle19},
LNCA~\cite{Quelle21} and XNCA~\cite{Quelle20}, which all used a
picture of effective sites and relied on NCA as the best available
implementation in those days. These three forms of effective site
theories aimed at a solution of the Anderson lattice model in the
context of the Heavy Fermion problem, and thus were based on a
particular local shell structure with well localized $f$-states
and extended $c$-states, with only the former being subject to a
local Coulomb repulsion. In most cases, transfer also was restricted
to the c-states only, which together makes possible a reduction from
the matrix formulation, envisaged in section 2 and shortly outlined
below, to a scalar formalism.

The three theories differed in the
way, in which the dynamics on general lattice sites influenced the
representative effective site considered: In ATA only the coherent
build up of quasiparticle bands from scattering by independent
Anderson-impurities on the lattice sites was taken into account thus
ignoring exhaustion problems and important renormalization effects.
LNCA used self-consistently modified local excitations at the
effective site; unlike XNCA, however, it introduced weight factors
for nonlocal processes in order to approximately factorize the
partition function into contributions from an unperturbed band and
independent effective sites. XNCA finally established the form of
self-consistency between effective site and surrounding medium which
becomes exact in the limit of infinite spatial dimension and which
nowadays is regarded as characteristic defining feature of the
DMFT.

In a local approach a lattice Hamiltonian
$H=\sum\limits_\nu{H_{0\nu}}+\sum\limits_{\nu\neq\nu'}V_{\nu,\nu'}$
is built up from local Hamiltonians $H_{0\nu}$ on lattice sites
$R_\nu$, each of the type $H_{0\ell}$ considered in section 2, and
nonlocal parts $V_{\nu,\nu'}$, which contain one-particle terms like
transfer or hybridization and possibly two-particle terms, i.e.\
nonlocal interactions between electrons on two different sites $\nu$
and $\nu'$. $V_{\nu,\nu'}$ is usually
expressed via the elementary creation and annihilation operators
$f^{(+)}_{\nu{m}\sigma}$ which define the local Fock-spaces, whereas
the diagonalization of $H_{0\nu}$ involves the "ionic states" as
described in section 2. In principle there is complete freedom in
defining what should be such a local subsystem, a single ionic
shell, or an ion or a molecular complex, or even a cluster of ions
or complexes. The need to diagonalize them, even when restricted to
e.g.\ the low energy regime, may however set limitations to the size
of what can be regarded as local. At the outcome $H_{0\nu}$ will be
a finite matrix in the space of local many body eigenstates
$|\nu\,{n_0}\,M\rangle$. A one-particle transfer term contained in
$V_{\nu,\nu'}$ will e.g.\ induce changes
$|\nu\,{n_0}\,M_1\rangle\rightarrow|\nu\,{n_0}\!-\!1\,M_2\rangle,|\nu'\,n'_0\,M'_1
\rangle\rightarrow|\nu'\,n_0'\!+\!1\,M_2'\rangle$.

A convenient way of
keeping the formalism simple is to work generally with the original
one particle quantum numbers $(m,\sigma)$ and to built up local
matrix Greensfunctions of the type
$\uul{G}_\nu(z)=(G_{m\sigma,m'\sigma'}(z))$ for one-particle
propagation, and corresponding higher ones; these matrices involve
information about the composition of the local many-body
eigenstates, e.g.\ via the coefficients of fractional occupancy. Thus
an overall ($N\times N$)-matrix formalism for lattice processes is
established with N being the number of local one-particle states
taken into account, and local excitations of single electrons are
translated via \eqref{Gleichung2} into the dynamics of many-body eigenstates.
Nonlocal interactions can be viewed as (simultaneous) two-particle
transfers and be handled in an analogous fashion.

The local approach treats the local dynamics exactly, as apparent in
the many-body eigenstates. In the presence of strong and dominating
local interaction matrix elements this seems a natural starting
point for a lattice theory. However, it turns to a disadvantage
when a perturbation expansion in terms of the nonlocal parts
$V_{\nu,\nu'}$ of the Hamiltonian is to be set up. An attempt to
factorize e.g.\ a general contribution to the partition function with
the help of Wick큦 theorem stops at an intermediate level: In case of
e.g.\ a pure one-particle nearest-neighbour transfer mechanism, time
ordered expectation values on products of creation and annihilation
operators belonging to sites, which knot together different
propagation paths of single particles, remain as unfactorized parts
of internally connected contributions $B_\lambda$. These also
contain matrix elements $t$ of $V_{\nu,\nu'}$, symmetry factors
$r_\lambda$, a sign $(-1)^{\chi_\lambda}$, site-summations and
time-integrations, as well as an indicator function $F$ assuming
values 0 or 1,
which realizes site-exclusions between all the $B_\lambda$ occuring via $n_\lambda=1$). 
Schematically:
\begin{eqnarray}
\frac{Z}{Z_0}=\sum\limits_{\{n_\lambda=0,1\}}\left(\prod\limits^\infty_{\lambda=1}
  B^{n_\lambda}_\lambda\right)F(\{n_\lambda\}),
\nonumber\\
B_\lambda=\frac{(-1)^{\chi_\lambda}}{r_\lambda}
\Big(
\prod\limits_{\mathrm{sites}\: \mu \:\mathrm{involved\:as\:knots}}
\int{d}\tau^{(1)}_\mu\int{d}\tau^{(1)'}_\mu\cdot\ldots
\nonumber\\
\phantom{\frac{(-1)^{\chi_\lambda}}{r_\lambda}\prod\limits_{\mathrm{sites}\:\mu\:\mathrm{involved\:as\:knots}}} 
\cdot \langle
T(f_{m^{(1)}_\mu\sigma^{(1)}_\mu}(\tau^{(1)}_\mu)\:f^+_{m^{(1)'}_{\mu}\sigma
^{(1)'}_{\mu}}(\tau^{(1)'}_\mu)\cdot\ldots)\rangle_0
\Big)
\nonumber\\
\cdot
\prod\limits_{
  \mathrm{chains}\:\mathcal{C}\:\mathrm{of\: transfers\:between\:knots}
}
\:\:\prod\limits_{\mathrm{intermediate\:sites}\:\nu\:\mathrm{along}
  \mathcal{C}}\int{d}\tau^{(1)}_{\nu}\int{d}\tau^{(1)'}_{\nu}\cdot\ldots
\nonumber\\
\phantom{\cdot\prod\limits_{{\mathrm{chains}\:\mathcal{C}\:\mathrm{of\: transfers\:between}}}}
\cdot{t}\cdot\langle
T(f_{m^{(1)}_{\nu}\sigma^{(1)}_{\nu}}(\tau^{(1)}_{\nu})
\cdot{f}^{+}_{m^{(1)'}_{\nu}\sigma^{(1)'}_{\nu}}(\tau^{(1)'}_{\nu}))\rangle_0\cdot\ldots\cdot{t}
\end{eqnarray}
\begin{figure}[t!]
  \begin{center}
    \includegraphics[width=10cm]{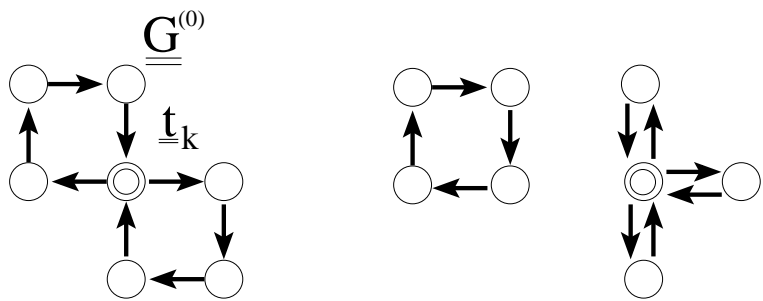}
  \end{center}
  \caption{This contribution to the partition function is a product of
three disconnected pieces, each of them containing single-particle
loops. In two of the pieces loops are glued together at sites
(nodes), which are marked with a double circle. These nodes give
rise to cumulant vertices for local two-and three-particle
interactions, respectively.}
  \label{Figure16}
\end{figure}
In general, local correlations between all of the excitations caused at
the nodes by intersite transfers remain. The emerging
picture is that of disconnected sets of loops over the lattice, each
set being internally glued together at certain sites with four or
more intersite transfer legs, see figure~\ref{Figure16} for a simple
example. Moreover, while keeping the topological structure of such a
graph, the position of sites involved cannot be summed freely over
the lattice, thus preventing a convenient momentum space
formulation. As a consequence, also a linked cluster theorem is not
available for Greensfunctions of the lattice problem, since a
partition function factor, the diagrams not linked to external
sources, cannot be factored out: the partition function can also not
be represented as an exponential of a sum of single connected graphs
thus preventing a straightforward extensitivity property of the
thermodynamic potential. 

A way out of this dilemma uses a
representation of higher order local time ordered expectation values
at nodes as a sum over products of successively smaller ones, which
in total cancel out except for the original highest term; when
appropriately grouped together they define a cumulant-expansion of
local $n$-particle Greenfunctions, containing a set $M_n$ of $n$ local
destruction
operators and a set $\widetilde{M}_n$ of creation operators:
\begin{eqnarray}
&G_n(M_n;\widetilde{M_n})=\hspace*{-1.5cm}\sum\limits_{
  \mathrm{all partitions}\:P\:\mathrm{of}\:M_n\:\mathrm{and}
  \:\widetilde{P}\:\mathrm{of}\:\widetilde{M}_n\:
  \mathrm{into\:subsets}\:\left({N}^{(q)}_{n_q},\widetilde{N}^{(q)}_{n_q}\right)
}
\hspace*{-1.5cm}(-1)^{\chi_p+\chi_{\widetilde{p}}}\prod\limits
_{q}G^c_{n_q}(N^{(q)}_{n_q},\widetilde{N}^{(q)}_{n_q}),\nonumber\\
&G^c_2\:(f_{m^{(1)}_\mu\sigma^{(1)}_\mu}(\tau^{(1)}_\mu),\:f_
{m^{(2)}_\mu\sigma^{(2)}_\mu}(\tau^{(2)}_\mu);\:f^+_{m^{(2)'}_
\mu\sigma^{(2)'}_\mu}(\tau^{(2)'}_\mu),\:f^+_{m^{(1)'}_
\mu\sigma^{(1)'}_\mu}(\tau^{(1)'}_\mu))\:=\nonumber\\
&G_2(1,2;2',1')-[G_1(1;1')G_1(2;2')-G_1(1;2')G_1(2;1')]\quad,\ldots\quad.
\end{eqnarray}
The right hand side of this expression can be viewed as containing
two contributions playing each a different role in the expansion:
The maximally decomposed terms, being products of only one-particle
Greensfunctions at this site, together just furnish the result which
would be obtained if Wick큦 theorem were valid for the local
dynamics. The rest of the terms represent all possible local
decompositions of the knot in the graph into products of independent
knots of lower order with together the same number of intersite legs
as the original expectation value; among them are possibly local
one-particle Greensfunctions but at least one of higher order, i.e.\
with more than two intersite legs. 

An evident approximation can be obtained as follows: 
Keeping only the maximally
decomposed terms, all loops of the diagram, which were glued
together originally, would be disconnected, the partition function
would become an exponential of the sum of all different loops, and a
free summation of all site positions over the lattice would leave
only contributions differing by the number of links. Likewise, e.g.\
the lattice-one-particle-Greensfunction would contain, after
dividing out the partition function, a sum over paths differing only
by the number of links, which can be summed as a geometric series
(in matrix space).

The results for one-particle Greensfunction and thermodynamic potential $F$
can be given explicitly as
\begin{eqnarray}
  &\uul{G}^{FT}_{\underline{k}}(z)=[\uul{G}^{(0)}(z)^{-1}-\uul{t}
  _{\underline{k}}]^{-1}\quad,\nonumber\\\label{Gleichung23}
  &F^{FT}-F^{(0)}=\frac{1}{\beta}\sum\limits_{i\omega_n}
  \sum\limits_{\underline{k}}(Tr[\ln\uul{G}^{(FT)}_{\underline{k}}
  (i\omega_n)]-Tr[\ln\uul{G}^{(0)}(i\omega_n)])
  \quad,
\end{eqnarray}
with $\uul{G}^{(0)}(z)$ the Greensfunction  from the known solution
of the isolated local subsystem.
The above solutions may be viewed as a very general form of the well known
Hubbard-I approximation~\cite{Quelle51}, to which they reduce when
the local subsystem is a simple s-shell and $V$ contains only
nearest-neighbour one-particle transfers i.e.\ in the case of the
Hubbard model.

The rest of the terms in the cumulant-decomposition
furnishes all possible combinations of one-particle scattering
events off this site, serving as nodes with two intersite legs in
simple transfer chains, and of higher order nodes with more
intersite legs; together they realize a restricted way of glueing
together the original connected set at this site. The contributions
resulting from all of these local cumulant decompositions in the set
can be re-interpreted in terms of a perturbation expansion with
respect to an infinite set of local n-particle cumulant
interactions, with n between 2 and infinity.

The "non-interacting"
starting point of this expansion is the aforementioned generalized
Hubbard-I-theory free of such interactions ("Free Theory"). A
perturbation expansion with the corresponding "free" propagators for
the particles and the set of all local cumulant vertices as
interaction terms along the conventional lines a la Feynman thus
faithfully produces all contributions to partition function and
lattice Greens functions and re-introduces the applicability of the
linked cluster theorem and all benefits connected with it into this
new form of the theory.

The prize paid for the conceptual progress described above for the
lattice problem. i.e.\ the applicability of conventional methods in
perturbation theory for the lattice aspects, lies in the large
number of vertices appearing and in their dynamical nature, i.e.\
their multiple time-dependencies. It should also be clear that the
perturbation series obtained in this way are based, although looking
conventional in a superficial way, on quite unconventional
definitions of connectedness and irreducibility: Whereas the
diagrams remaining i.e.\ for a one-particle-Greensfunction are
linked to the external sources and consist of one connected piece
(we exclude anomalous terms here), being glued together via local
cumulant vertices, they would in most cases fall apart into several
unconnected pieces, only one of them bearing the two external links,
when the cumulants are made explicit in terms of the original
expectation values. 

The pieces without links remaining in this case
as factors are partially due to the partition function in the
denominator of the original expression and must consistently be
included for a proper renormalization of e.g.\ excitation spectra. It
is therefore not a trivial problem to define consistent
approximations in cumulant perturbation theory. 

A straightforward
evaluation of cumulant vertices even in low orders requires some
effort but can be managed e.g.\ with methods of direct perturbation
theory~\cite{Quelle52}. The somewhat lengthy expression of the local
two-particle cumulant has been published before~\cite{Quelle53} and
served as basis for calculations of one-particle properties of the
Hubbard model along the lines of Hartree-like
expansions~\cite{Quelle54}, also in a self-consistent
fashion~\cite{Quelle52}. It should be clear and can in fact be
proven~\cite{Quelle52}, that these approximations are not able to
capture the interesting many body aspects in the low temperature
regime for large values of the local Coulomb repulsion $U$, whereas
leading effects of band-splitting, which are present in principle
already in the Hubbard-I approximation, and of band-deformation can
be captured. What is needed for a proper approximation of strong
correlation effects are summations of infinitely many processes with
cumulant vertices up to infinite order to cope with the problem of
long-time decay of correlations and of the infrared problems
connected with them.

The local starting point of the cumulant expansion and its
formulation in real space open the possibility of a local  infinite
order resummation, which would be more difficult to recognize after
unrestricted Fourier-transformation. In the latter
$\underline{k}$-space version of cumulant perturbation theory, due
to the unrestricted site-summations, processes taking place on the
same site in a diagram, cannot be identified anymore and may appear
completely uncorrelated. 

Moreover, also individual parts of the
partition function in the original denominator connected to the same
site and necessary for a proper local normalization are hidden in
the cumulant and cannot readily be identified. A proper conserving
approximation of infinite order should unite such pieces of local
processes in a consistent way. This furnishes the guideline for a
"locally complete approximation": Collect all those diagrammatic
contributions to the 1-P-Greensfunction of cumulant perturbation
theory in real space, where both external legs belong to the same
site, not regarding whether they belong to the same or to different
vertices situated at this site.

\begin{figure}[t!]
  \begin{center}
    \includegraphics[width=12cm]{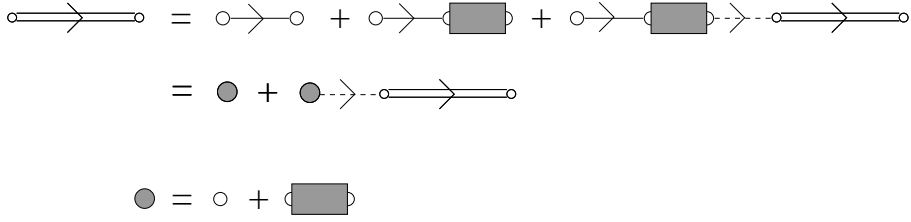}
  \end{center}
  \caption{Different forms of the Dyson-equation, for the case of
one-particle transfer only, in cumulant perturbation theory. Care
has to be taken in specifying nonlocal connections, which is done
here via the broken transfer lines made explicit. Single unbroken
lines relate to the propagators in Free Theory, the hatched
rectangular box is one-particle irreducible in cumulant (and not in
regular) perturbation theory.}
  \label{Figure17}
\end{figure}
The formulation of a Dyson-equation for the
1-particle-Greensfunction in cumulant perturbation theory is
straightforward, but needs some care in distinguishing local and
nonlocal parts of the propagation process: certain
intersite-transfers have to be made explicit. In a straightforward
way one-particle-irreducible pieces can be identified, which either
are linked to external sources or to transfers. It is useful to
amputate local factors $\uul{G}^{(0)}(z)$ (the Greens functions
of isolated local subsystems) at their two ends, thus defining
irreducible cumulant selfenergies
$\uul{\Sigma}^{(amp)}_{\underline{k}}(z)$. Since the two ends can
be situated at two different sites, they will generally be
$\underline{k}$-dependent after Fourier-transformation. Two
equivalent forms of the Dyson-equation are visualized in
figure~\ref{Figure17}; their algebraic form is:
\begin{eqnarray}
  \label{Gleichung24}
  \uul{G}_{\underline{k}}(z)&=\uul{G}_{\underline{k}}^{(FT)}(z)+\uul{G}_{\underline{k}}
  ^{(FT)}(z)\uul{\Sigma}_{\underline{k}}^{(amp)}(z)\uul{G}^{(0)}(z)+\uul{G}_{\underline{k}}
  ^{(FT)}(z)\uul{\Sigma}_{\underline{k}}^{(amp)}(z)\uul{G}^{(0)}(z)\uul{t}_{\underline{k}}\uul{G}
  _{\underline{k}}(z)
  \nonumber  \\
  &=\uul{\widetilde{G}}_{\underline{k}}(z)+\uul{\widetilde{G}}_{\underline{k}}(z)
  \uul{t}_{\underline{k}}\uul{G}_{\underline{k}}(z)
  \quad \Rightarrow \quad\uul{G}_{\underline{k}}(z)=\left[\uul{\tilde G}_{\underline{k}}(z)^{-1}-\uul{t}_{\underline{k}} \right]^{-1}
  \quad,
\end{eqnarray}
with
\begin{eqnarray}
  \uul{G}_{\underline{k}}^{(FT)}(z)&=\left[\uul{G}^{(0)}(z)^{-1}-\uul{t}
    _{\underline{k}}\right]^{-1}
  \equiv \left[z -\uul{\Sigma}^{(0)}(z)-\uul{t}_{\underline{k}}\right]^{-1}
  \\
  \uul{\widetilde{G}}_{\underline{k}}(z)&=\uul{G}^{(0)}(z)+
  \uul{G}^{(0)}(z)\uul{\Sigma}^{(amp)}_{\underline{k}}(z)\uul{G}^{(0)} (z)
  \quad,
\end{eqnarray}
\begin{figure}[t!]
  \begin{center}
    \includegraphics[width=10cm]{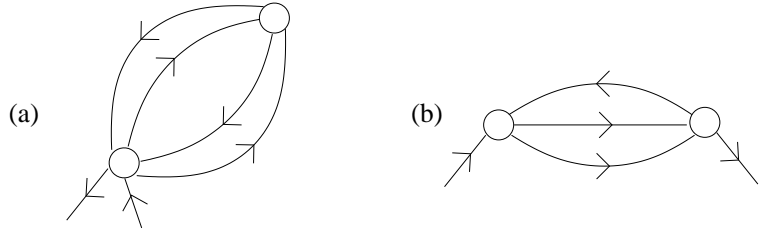}
  \end{center}
  \caption{Contributions neglected in the selfenergy-matrix in the
``locally complete'' (LC) approximation, one-particle transfer only.
Part (a) shows a process, in which two loops, based at the same site,
are correlated at a different site by a 2-particle cumulant vertex.
Part (b) shows a process, in which the external particle is
extracted at a site different from where it was injected.}
  \label{Figure18}
\end{figure}
where $\uul{\Sigma}^{(0)}(z)$ is known from the solution of the local subsystem.
The irreducible cumulant selfenergy $ \uul{\Sigma}^{(amp)}_{\underline{k}}(z)$ defined  above
must not be confused  with the standard selfenergy $\uul{\Sigma}^{(st)}_{\underline{k}}(z)$ defined via
\begin{eqnarray}
  \label{eq:standardSE}
  \uul{G}_{\underline{k}}(z)&=\left[z-\uul{\Sigma}^{(st)}_{\underline{k}}(z)-\uul{t}_{\underline{k}}\right]^{-1}
  \quad.
\end{eqnarray}
The connection between them, which can be expressed as
\begin{eqnarray}
  \label{eq:SEs}
  \uul{\Sigma}^{(st)}_{\underline{k}}(z)&=
  \uul{\Sigma}^{(0)}(z)
  +\uul{\Sigma}^{(amp)}_{\underline{k}}(z)\left[1+\uul{G}^{(0)}(z)\uul{\Sigma}^{(amp)}_{\underline{k}}(z)\right]^{-1}
  \quad,
\end{eqnarray}
sheds some light on the possible momentum dependence of
the selfenergy. We expect that this may contribute to the question
of how to incorporate nonlocal correlations into lattice theories (see ~\cite{Quelle27,Quelle57a}
and refercences therein).

In a local approximation, the two external links of
$\uul{\Sigma}^{(amp)}_{\underline{k}}(z)$ are restricted to be
situated at the same site, which eliminates the
$\underline{k}$-dependence from $\uul{\Sigma}^{(amp)}(z)$. 
This type of approximation is
quite in the tradition of the early effective-site theories
mentioned above. It brings formal advantages but involves
shortcomings concerning the neglect of certain nonlocal
correlations. 

Still it does not lead to an easily tractable
calculational scheme, which is due to classes of remaining nonlocal
correlations in $\uul{\Sigma}^{(amp)}(z)$ as indicated in
figure~\ref{Figure18}(a); part (b) of this figure on the other hand
shows correlations not included in the local form of
$\uul{\Sigma}^{(amp)}(z)$. Eliminating diagrams like the one
shown in figure~\ref{Figure18}(a) leads to a restricted form
$\uul{\Sigma}^{(lc)}(z)$ of $\uul{\Sigma}^{(amp)}(z)$, which
is at least "locally complete" and may be characterized as follows:
In a diagram contributing to $\uul{\Sigma}^{(lc)}(z)$ 
cumulant vertices of order $n\geq{2}$ (four or more external links)
are forbidden to appear, since they would correlate two or more 
1-particle loops based at the same site. Therefore 1-particle loops
generally will contain
insertions with two external links (and no more), which are of the
same class as those contributing to $\uul{\Sigma}^{(lc)}(z)$; in
this way a complete hierarchical structure of independent loops
remains.

Amazingly, the locally complete approximation thus defined
can quite generally be brought into a form suited for a
straightforward solution, in principle without further
approximations or restrictions. It turns out to be equivalent to
general formulations of the XNCA- or the
DMFT-methods~\cite{Quelle20,Quelle52}.
The key to a simpler formulation of the locally complete
approximation of cumulant perturbation theory lies in the reduction
of all nonlocal topological elements in a diagram to independent
one-particle loops as described above. This makes it possible to
trace back the $\underline{k}$-summed form of
~(\ref{Gleichung24}), i.e.\
\begin{eqnarray}
\uul{G}(z)\equiv\frac{1}{N}\sum\limits_k\uul{G}_{\underline{k}}(z)=
\frac{1}{N}\sum\limits_{\underline{k}}\left[\uul{\widetilde{G}}(z)^{-1}-
\uul{t}_{\underline{k}}\right]^{-1},\label{Gleichung25}
\end{eqnarray}
to the irreducible part $\uul{\widetilde{T}}(z)$ of the
(unrestricted) loop-propagator
\begin{eqnarray}
\uul{T}(z)=\frac{1}{N}\sum\limits_k\uul{T}_{\underline{k}}(z)\quad,\quad\uul{T}_{\underline{k}}(z)=
\uul{t}_{\underline{k}}\uul{G}_{\underline{k}}(z)\uul{t}_{\underline{k}}
\label{Gleichung26}
\end{eqnarray}

It turns out that $\uul{G}(z)$, as the local
one-particle-Greensfunction (i.e.\ both external sources are at the
same site), can be constructed as a functional
$\uul{G}[z;\uul{\varrho}_{\widetilde{T}}(\omega)]$ of the
irreducible loop-spectrum
$\uul{\varrho}_{\widetilde{T}}(\omega)=-\frac{1}{\pi}Im\widetilde
{\uul{T}}(\omega+i\delta)$, and $\widetilde
{\uul{G}}(\uul{G},\widetilde {\uul{T}})$ has a simple
expression via $\uul{G}$ and $\uul{\widetilde{T}}$. Therefore
(\ref{Gleichung26}) reduces to an implicit equation for
$\uul{\widetilde{T}}(z)$ and consequently also furnishes
solutions for
$\uul{\widetilde{G}}(z)\:,\:\uul{G}_{\underline{k}}(z)$ and
$\uul{G}(z)$. 

The original loop propagator $\uul{T}(z)$
embodies the possibility that the loop connects to its basic site at
several intermediate instances due to the unrestricted
site-summations contributing to $\uul{G}_{\underline{k}}(z)$ in
cumulant perturbation theory. The whole loop may be viewed as a
repetition of irreducible pieces, i.e.\ loops without such
intermediate connections to its basic site. The relation between
both loop propagators takes the form of a Dyson-equation
\begin{eqnarray}
\uul{T}(z)=\uul{\widetilde{T}}(z)+\uul{T}(z)\uul{\widetilde{G}}(z)\uul{\widetilde{T}}(z)
\:\Rightarrow\:\uul{\widetilde{G}}(z)=\uul{\widetilde{T}}(z)^{-1}-\uul{T}(z)^{-1}\:,
\label{Gleichung27}
\end{eqnarray}
which was first recognized in~\cite{Quelle55}.
~(\ref{Gleichung27}) generalizes earlier attempts for
implementing a "site-exclusion principle" for the propagation of
quasiparticles under the influence of strong local correlations
~\cite{Quelle3}.

It is easy to visualize the effect of loops on the dynamics of the
basic site of $\uul{\Sigma}^{amp}(z)$, i.e.\ the one bearing the
two external links, when e.g.\ the picture suggested by direct
perturbation theory is used. Apart from the fact, that a projection
of $\uul{\varrho}_{\widetilde{T}}(\omega)$ onto a local 1-P-state
is used in all formulas instead of the spectrum
$\varrho^{(0)}_{c\sigma}(\omega)$ of band electrons, the way to
calculate $\uul{G}(z)$ is the same as in the impurity problem
considered in section 2. 

Insofar, an effective impurity is
constructed, and the spectrum of the irreducible loop-propagator may
be viewed as a matrix of real, frequency-dependent effective
external fields, constituting a "bath" or a "dynamical mean field".
Conclusions about $\uul{\widetilde{G}}(z)$ are to be traced back
to the definition
$\uul{\widetilde{G}}(z)=\uul{G}^{(0)}(z)+\Delta\uul{\widetilde{G}}(z)$,
$\Delta\uul{\widetilde{G}}(z)=\uul{G}^{(0)}{(z)}\uul{\Sigma}^{amp}(z)\uul{G}^{(0)}(z)$
in ~(\ref{Gleichung24}), reproduced here for the locally complete
approximation.

At first glance it may seem as if the diagrams
contributing to $\uul{\widetilde{G}}(z)$ in cumulant perturbation
theory via $\uul{\Sigma}^{amp}(z)$ would just reproduce the
original contributions of the effective impurity problem. However, this 
cannot be true, since the introduction of cumulants also serves
the purpose of removing restrictions from site-summations, giving
$\uul{G}_{\underline{k}}(z)$ the simple form used in
~(\ref{Gleichung25}). $\uul{\widetilde{G}}(z)$ thus contains
compensation terms for loop-contributions produced by
$\uul{t}_{\underline{k}}$ in the denominator, i.e.\ in the
corresponding geometric series with local parts
$\uul{\widetilde{G}}(z)$ and links $\uul{t}_{\underline{k}}$.
One possible way of uncovering the relation between $\uul{G}(z)$
and $\uul{\widetilde{G}}(z)$ consists in formulating the
difference between both quantities just as the contribution for
loops to be compensated, i.e.\ \footnote{ 
  An alternative statement of this relation is 
  \begin{equation*}
    \uul{\widetilde{G}}(z)\uul{T}(z)=\uul{G}(z)\uul{\widetilde{T}}(z)\:,
  \end{equation*}
  which is realized by analyzing the expansions of $\uul{G}(z)$ and 
  $\uul{T}(z)$ in terms of irreducible loops $\uul{\widetilde{T}}(z)$.
}
\begin{eqnarray}
\uul{\widetilde{G}}(z)-\uul{G}(z)=-\uul{\widetilde{G}}(z)\uul{T}(z)\uul{\widetilde{G}}(z).
\label{Gleichung28}
\end{eqnarray}
When ~(\ref{Gleichung27}) is inserted into
~(\ref{Gleichung28}) to eliminate $\uul{T}(z)$ in favour of
$\uul{\widetilde{T}}(z)$, one obtains
\begin{eqnarray}
\uul{\widetilde{G}}(z)=\Big[\uul{G}(z)^{-1}+\uul{\widetilde{T}}(z)\Big]^{-1}\equiv
\uul{\widetilde{G}}(\uul{G},\uul{\widetilde{T}})(z),
\label{Gleichung29}
\end{eqnarray}
thus completing the reduction of ~(\ref{Gleichung25}) to an
implicit equation for $\uul{\widetilde{T}}$ as envisaged above.

It should finally be remarked that the contributions in cumulant
perturbation theory to the quantities considered here, which may be
classified as connected pieces not linked to the external sources in
the original picture, are absorbed in the proper normalization of
spectra and 1-particle-Greensfunctions; they originate from a
division by the partition function as explained above. Using a
consistent locally complete summation of these contributions, they
become absorbed in the partition function of the effective site
problem. In direct perturbation theory, for example, this is taken
into account by properly normalized defect propagators.

\medskip

\subsection{Periodic Anderson Model: Part I}

The formal development outlined above leads to a result, which
constitutes a matrix generalization of DMFT. The
original concern of LNCA and XNCA was the physics of the Anderson
lattice model,
\begin{equation}
  \label{eq:LAMH}
  \hat H =\sum_{\sigma,i} \left(
    \epsilon_{\ell} \:\hat{f}^\dagger_{i\sigma}\hat{f}_{i\sigma}
    +\frac{U}{2}\,
    \hat{n}^f_{i\sigma}\hat{n}^f_{i\bar{\sigma}}
  \right)
  +\sum_{\k,\sigma}\epsilon_{\k}\,\hat{c}^\dagger_{\k\sigma}\hat{c}_{\k\sigma}
  +\sum_{\k,\sigma} \left(V_{\k}\,\hat{c}^\dagger_{\k\sigma}\hat{f}_{\k \sigma}+h.c. \right) 
\end{equation}
 and it should shortly be explained, how a scalar form
of the equations is achieved for this case. The local subsystem here
involves for the simplest case a basis of four one-particle states,
the two $f$-states with spin up and down subject to the
Coulomb-repulsion $U$ and two $c$-states which do not interact
with each other or with the $f$-states. All
matrices considered above are four by four, with $\uul{t}$ being
spin-diagonal and transferring only $c$- and $f$-electrons to
nearest-neighbour $c$-states with respective matrix elements $t$ and
$V$. 

Although unphysical in most cases, a purely local hybridization
$V$ is often used for simplicity; it can be treated in close analogy
to the nearest-neighbour case for the reason explained in the
following. Since the $c$-electrons remain noninteracting, Wicks
theorem can be used for them. Consequently, no cumulant vertices of
order $n\geq2$ exist with links to $c$-Greensfunctions.
$\uul{G}^{(0)}(z)$ is block-diagonal with a diagonal $c$-block
and a $f$-block, and likewise is $\uul{G}^{(0)}(z)^{-1}$, which
is used in the amputation of vertices. 

The block-structure mentioned
is $2\times2$ with respect to spin; without magnetic splitting and
with spin-preserving transfer and hybridization one may fix a
spin-direction $\sigma$ and treat these blocks as scalars. As a
consequence, $\uul{\Sigma}^{amp}(z)$ and also
$\Delta\uul{\widetilde{G}}(z)$ have nonzero matrix elements only
in the diagonal, i.e.\ only $\widetilde{G}_{ff\sigma}$ and
$\widetilde{G}_{cc\sigma}(z)=\widetilde{G}^{0}_{cc\sigma}(z)$ enter
the calculation according to the definition in
~(\ref{Gleichung24}). Nondiagonal elements come into play only
via $\uul{\widetilde{T}}(z)$, since propagation along a loop can
mix $c$- and $f$-states. ~(\ref{Gleichung29}) now gives after
matrix-inversions:
\begin{eqnarray}
\widetilde{T}_{ff\sigma}(z)&=\widetilde{G}_{ff\sigma}(z)^{-1}-\frac{G_{cc\sigma}(z)}{N_\sigma{(z)}}\quad,
\quad\widetilde{T}_{cc\sigma}(z)=\widetilde{G}_{cc\sigma}(z)^{-1}-\frac{G_{ff\sigma}(z)}{N_\sigma{(z)}},\nonumber\\
\widetilde{T}_{cf\sigma}(z)&=\widetilde{T}_{fc\sigma}(z)=\frac{G_{cf\sigma}(z)}{N_\sigma{(z)}}\quad,\quad
N_\sigma(z)=G_{ff\sigma}(z)G_{cc\sigma}(z)-G_{cf\sigma}(z)^2.
\label{Gleichung30}
\end{eqnarray}
These matrix elements all would have to be used if the four states
were locally correlated. However, in the simple form of the Anderson model
the local $c$-states do not directly influence the
$f$-state dynamics. Therefore one can reduce the local problem to
one of $f$-states by combining site-irreducible loops in a way that
only a restricted form of irreducibility with respect to the
$f$-states on the basic site is realized, i.e.\ the $c$-level on
this site is treated like a different site. 

The combined
loop-propagator connects local $f$-states and obeys a generalized
Dyson-equation of the form
$\widetilde{T}^{(red)}_{ff\sigma}(z)=\widetilde{T}_{ff}+\widetilde{T}_{fc}\widetilde{G}_{cc}\widetilde{T}_{cf}
+\widetilde{T}_{fc}\widetilde{G}_{cc}\widetilde{T}_{cc}\widetilde{G}_{cc}\widetilde{T}_{cf}+\ldots$
and hence may be summed up to:
\begin{eqnarray}
\widetilde{T}^{(red)}_{ff\sigma}(z)=\widetilde{T}_{ff\sigma}(z)+\frac{\widetilde{T}_{cf\sigma}(z)^2
\widetilde{G}_{cc\sigma}(z)}{1-\widetilde{G}_{cc\sigma}(z)\widetilde{T}_{cc\sigma}(z)}\:.
\label{Gleichung31}
\end{eqnarray}
If now the quantities $\widetilde{T}$ on the r.h.s. are replaced via
~(\ref{Gleichung30}), one obtains the scalar equivalent to
~(\ref{Gleichung29}):
\begin{eqnarray}
\widetilde{T}^{(red)}_{ff\sigma}(z)=\widetilde{G}_{ff\sigma}(z)^{-1}-G_{ff\sigma}(z)^{-1}\:.
\label{Gleichung32}
\end{eqnarray}
As is clear from this construction, the spectrum of
$\widetilde{T}^{(red)}$ is to be used in the effective site problem,
i.e.\
$G_{ff\sigma}(z)=G_{ff\sigma}\left[z;\varrho_{\widetilde{T}^{(red)}_{ff\sigma}}(\omega)\right]$.

Also the inverse of the last term in ~(\ref{Gleichung32}), i.e.\
$G_{ff\sigma}(z)$, has to be calculated via ~(\ref{Gleichung25})
using a matrix-inversion, which involves the non diagonal
transfer-matrix
$\uul{t}_{\underline{k}}=t_{\underline{k}}\:\uul{1}+V_{\underline{k}}\:\uul{\sigma}_x$.
After explicitly formulating this step the reduction from a matrix-
to a scalar form of the theory for the Anderson-lattice model is
completed. 

Whereas there might exist easier ways of setting up the
XNCA/DMFT-self-consistency cycle for this model~\cite{Quelle20,QuellePAM}, the above
argumentation generally demonstrates the connection between the
universal matrix-formulation and possible reduced schemes. If, for
example, electrons in $c$-states would interact locally (but again
not with those on $f$-states), the matrix problem would reduce to
two scalar problems of type (28), which would be coupled only via
the lattice-summation, i.e.\ via the $\underline{k}$-sums in
~(\ref{Gleichung25})). This also points to a possible treatment
of more general models with inequivalent sites.

Finally, since
formally the local $c$-state acts like a different site on its
$f$-states, a local hybridization can be treated in the same way as
outlined above: The first and last transfer step, represented in
~(\ref{Gleichung26}) by factors $\uul{t}_{\underline{k}}$, now
carries a $\underline{k}$-independent matrix element $V$. This only
enters the calculation in a modified
$\uul{t}_{\underline{k}}=t_{\underline{k}}\:\uul{1}+V\uul{\sigma}_x$
to be used in ~(\ref{Gleichung25}).

\medskip

The Anderson-lattice model furnishes a good testing ground for the
quality of impurity-solvers because of the particular impact of
coherence in the half-filled case. In the symmetric situation of the
simple version without orbital degeneracy considered above, with two
electrons per site, the Luttinger-theorem predicts a Fermi surface
filling the whole first Brillouin zone, which should lead to the
formation of an excitation gap due to Bragg scattering in the
quasiparticle-DOS for temperature $T$ approaching
zero~\cite{Quelle57,Quelle58}. This signature of onsetting coherence
is hard to recover in approximations, since it requires a pronounced
structure in the selfenergy. 

On the one hand the increasing lifetime
of quasiparticles with $T\rightarrow0$ near the Fermi level requires
the (near) cancellation of the term $\Delta_A$ (see last section) in
the local selfenergy by the builtup of scattering during propagation
along loops through the lattice (see above), and on the other hand
formation of a gap should go along with a narrow peak in
Im$\Sigma$ signalling strongly increased resonant
scattering. This implies a fine balance between different
contributions, which is easily destroyed by inconsistent
approximations.

\subsection{Hubbard Model}
 
In order to elucidate the effect of increasing lifetime as $T\to 0$
one may consider the Hubbard model, 
\begin{equation}
  \label{eq:HubH}
  \hat H =\sum_{\sigma,\k} 
  \epsilon_{\k} \:\hat{c}^\dagger_{\k\sigma}\hat{c}_{\k\sigma}
  +\frac{U}{2}\sum_{\sigma,i}
  \hat{n}^c_{i\sigma}\hat{n}^c_{i\bar{\sigma}}
\end{equation}
where at half-filling and zero
temperature scattering should be absent near $\omega=0$ in the
Fermi-liquid phase, but which should not develop a coherence gap:
Since according to Luttinger큦 theorem the Fermi surface now lies
well inside the first Brillouin zone no Bragg scattering should be
effective there. 

Figure~\ref{Figure19}(a) shows the corresponding local
DOS, obtained within different approximations, two of them within
the locally complete scheme (DMFT) using SNCA and ENCA as impurity
solvers, respectively. 

%
\begin{figure}[p]
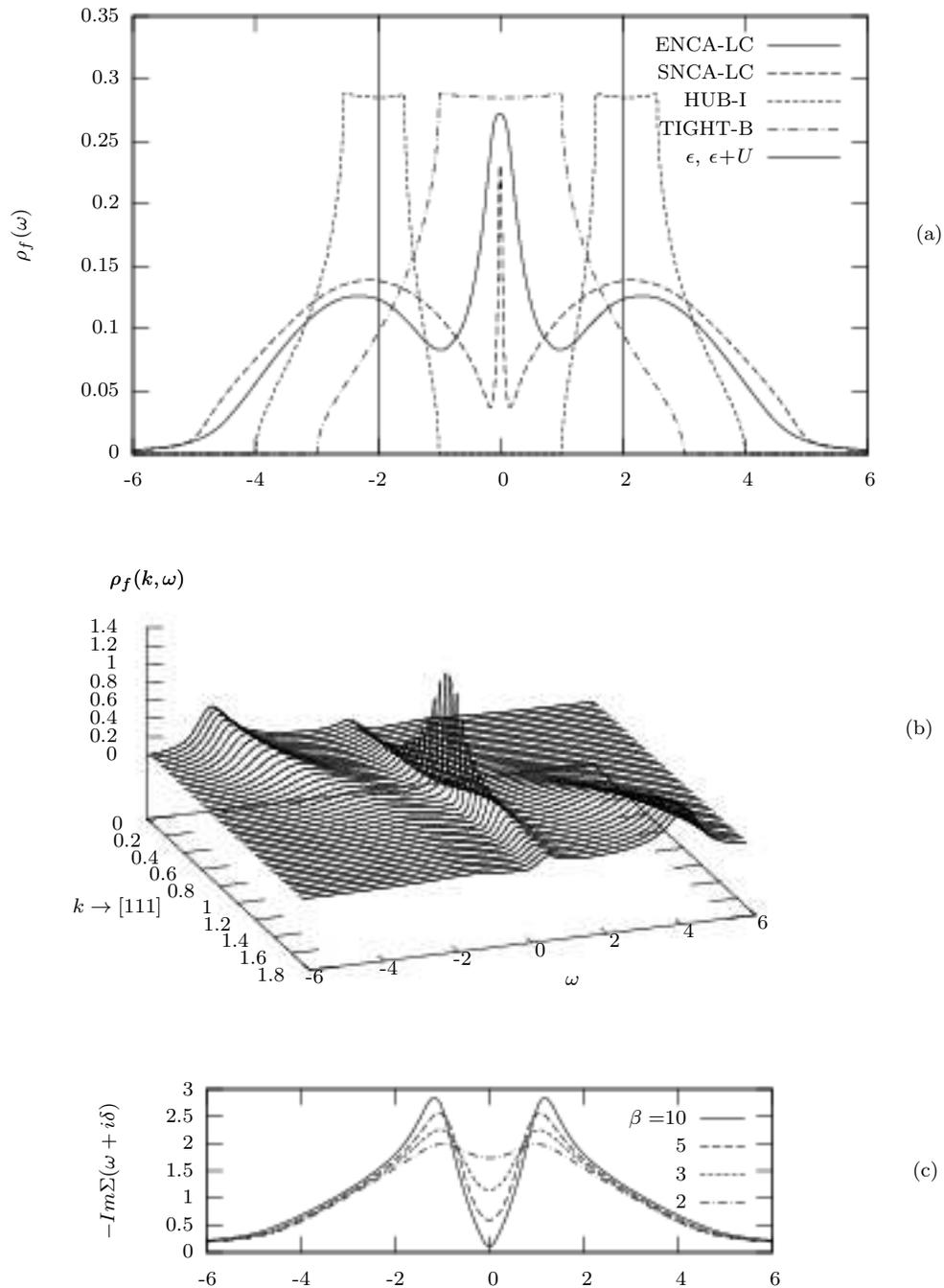

  \begin{center}
    {  \scriptsize
      \hspace*{-1.5cm}\input{Fig_19a.tex}\\[-4mm]
      \hspace*{-1.5cm}\input{Fig_19b.tex}
      \hspace*{-1.5cm}\input{Fig_19c.tex}
      \vspace*{-1cm}
  }
  \end{center}
  \caption{One- particle excitation spectrum of the Hubbard model,
    calculated within XNCA/DMFT, using ENCA as impurity solver, for the
    half-filled case with nearest-neigh bour hopping in a 3d-sc lattice.
    Parameter values $\epsilon=-2$, $U=4$ and $\beta$ not too large
    favour a metallic phase with a Fermi liquid. In part (a) the local
    DOS is compared for different approximation schemes, also including
    Hartree Fock (resulting in a tight-binding band of width 6) and the
    Hubbard-I approximation. Beside ENCA with $\beta=10$ also SNCA with
    $\beta=100$ is used as impurity solver. Part (b) shows the
    $\underline{k}$-resolved excitation spectrum along the
    [111]-direction, from which a quasiparticle bandstructure may be
    derived, within XNCA/DMFT-ENCA. The imaginary part of the selfenergy
    (absolute value) in part (c) visualizes the formation of the Fermi
    liquid with decreasing temperature $(k_B)T=\beta^{-1}$. }
  \label{Figure19}
\end{figure}
A reduction process from the matrix-formalism analog to the 
Anderson lattice model
is not necessary for the Hubbard model,
since the latter is of scalar type from the outset.

The other approximations are Hartree-Fock,
effectively meaning $U=0$ in the half-filled case, and Hubbard-I
("Free Theory"). The last two cases show no temperature dependence,
whereas $\beta=100$ and $\beta=10$ are chosen for SNCA and ENCA in
order to produce comparable heights of the many-body resonance,
which should reach the Hartree-Fock value at $T=0$.

In a
$\underline{k}$-resolution this DOS produces the
quasiparticle-bandstructure, as shown for the ENCA-calculation in
figure~\ref{Figure19}(b) along the [111]-direction of the simple cubic
Brillouin zone. At high excitation energies the two split bands,
which in Hubbard-I approximation contain sharp resonances with
reduced spectral weight, are so much washed out that a
concept of band electrons can hardly be justified here; this is in accord
with the conclusion in the last section about local scattering near
an impurity. The narrow band of pronounced $QP$-resonances around
$\omega=\mu=0$ shows no splitting and gap-formation. The
corresponding decrease of scattering is shown in
figure~\ref{Figure19}(c): In lowering the temperature from $\beta=2$
to $\beta=10$ the imaginary part of the self-energy forms a steep
and nearly quadratic minimum as a sign of Fermi-liquid formation.

\begin{figure}[hp!]
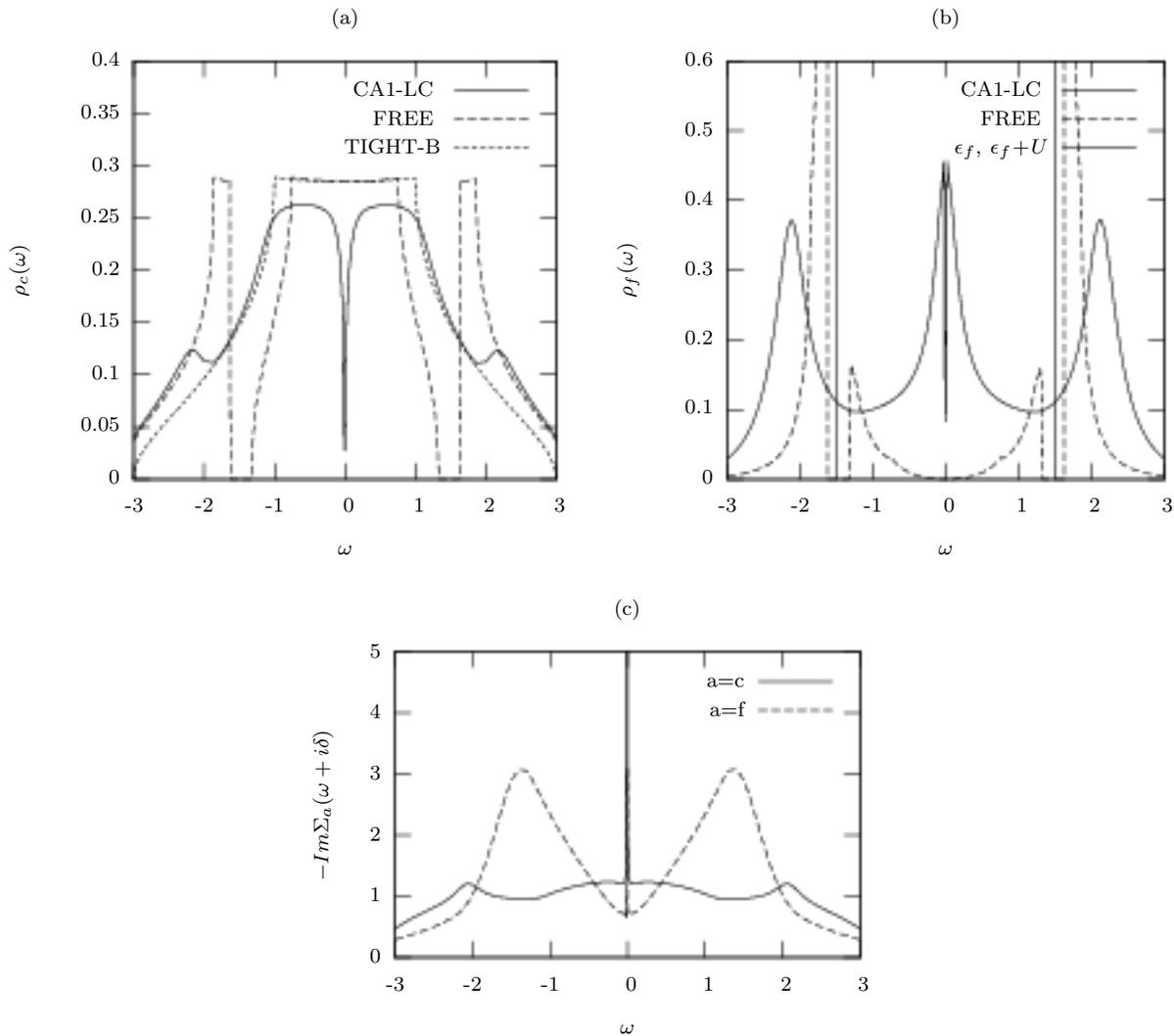

  \begin{center}
    {  \scriptsize
      \hspace*{-1.5cm}\input{Fig_20a.tex}\input{Fig_20b.tex}
      \hspace*{-1.5cm}\input{Fig_20c.tex}
    }
  \end{center}
  \caption{One-particle excitation spectra of band ($c$-) electrons and
    of local ($f$-) electrons for an Anderson-lattice model, calculated
    within XNCA, using CA1 as impurity solver, for the half-filled case
    (two electrons per site), with a tight-binding c-band of width 6 in
    a 3d-sc lattice and local hybridization. Parameters are
    $\epsilon_\ell=-1.5,\:U=3,\:\beta=15$ and $\Delta_A\equiv\pi V^2
    \varrho^{(0)}_{c\sigma}(0)=0.3$. Parts (a) and (b)  show the local
    density of c-and f-electrons, respectively, both in comparison with
    a Hartree-Fock result and a calculation within ``Free Theory''. Part
    (c) contains the imaginary part of the ``local selfenergies''
    (absolute value) for c-and f-electrons. The spikes seen in the
    middle of the gap region can cause numerical problems.
  }
  \label{Figure20a}
\end{figure}
\begin{figure}[hp!]
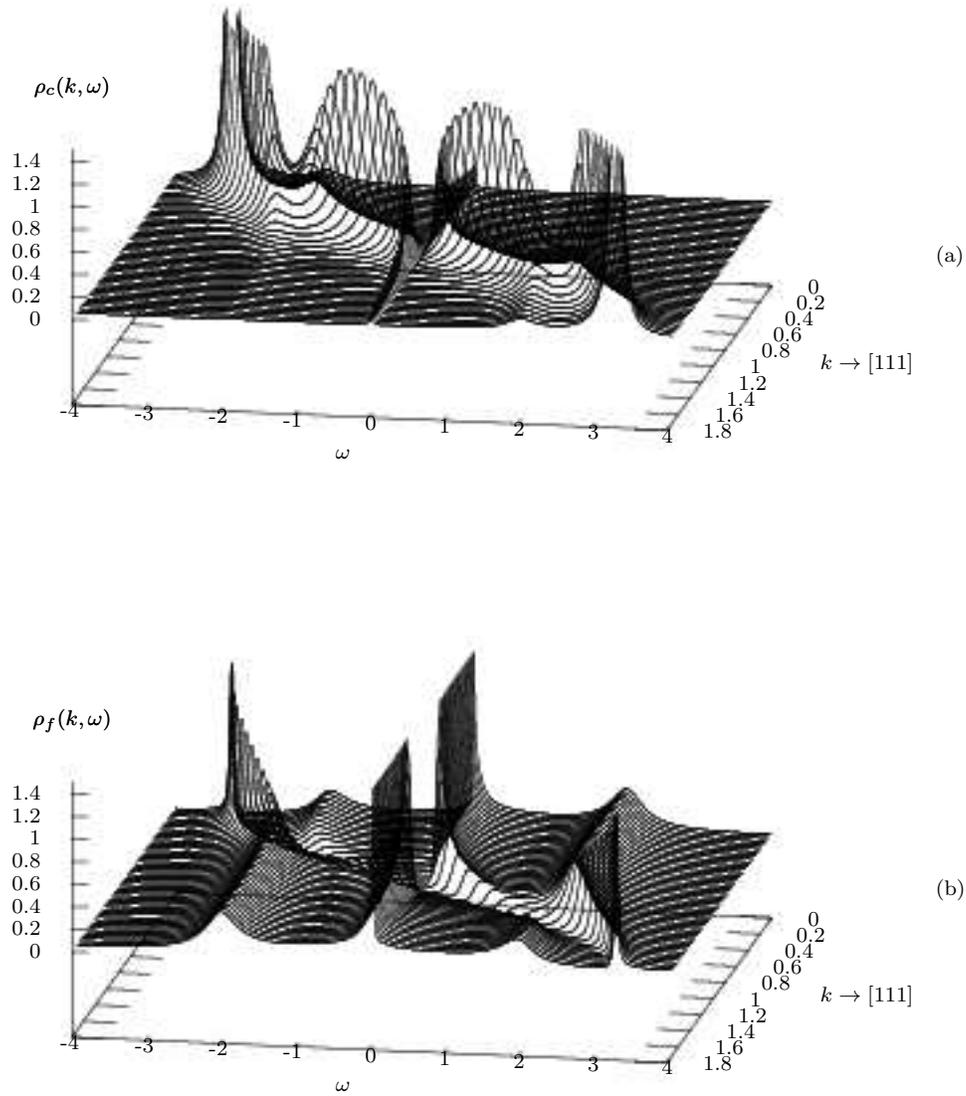

  \begin{center}
       {  \scriptsize
         \hspace*{-1.5cm}\input{Fig_20d.tex}
         \hspace*{-1.5cm}\input{Fig_20e.tex}
       }
  \end{center}
  \caption{
    The $\underline{k}$-resolved excitation spectra of c-and
    f-electrons, respectively, for the same parameters as in \ref{Figure20a} 
    are drawn along the [111]-direction. They
    demonstrate, like parts  \ref{Figure20a}(a) and (b), the formation of hybridization
    pseudogaps, smeared by lifetime effects, in the high-energy region
    and of a narrow and complete coherence gap at the Fermi energy. }
  \label{Figure20b}
\end{figure}

Quite generally it can be stated, that the ENCA impurity solver for the Hubbard 
model produces good results and 
accomplishes an acceptable tradeoff between accuracy and numerical effort. 
For not too low temperatures it thus represents a reliable  and usable alternative to impurity 
solvers like NRG, QMC, MPT, \dots known from the literature (see 
for example~\cite{QuelleHM,Quelle22}).

\subsection{Periodic Anderson Model: Part II}

The Anderson-lattice model behaves differently than the Hubbard model,
as shown in figure~\ref{Figure20a} and~\ref{Figure20b}. 
Parts~\ref{Figure20a}(a) and (b) contain the local one-particle 
excitation spectra for $c$- and
$f$-electrons, respectively, each calculated with three
approximations of increasing complexity. The tight-binding approach
for the band (c-) states treats all interactions on the mean-field
level and furnishes the connected curve known from the Hubbard model
with its edge-like van-Hove-singularities at $\omega=\pm{1}$ and
square-root band edges at $\omega=\pm{3}$. 

Hybridization with the
correlated local $f$-states at one-particle energies
$\omega=\epsilon_\ell=-1$ and $\omega=\epsilon_\ell+U=+1$ produces
the gaps visible in the result of the Free Theory, which furnishes
three disconnected bands with variable spectral weights. The gaps
are somewhat displaced by level repulsion, which is an effect of
hybridization, too. Interactions, much better taken into account in
the locally complete approximation using a CA1-impurity solver, wash
out the two gaps and produce a repulsion of $c$-weight away from the
Fermi level $\omega=0$ as a consequence of the formation of the
many-body resonance with predominant $f$-character. This resonance
is clearly seen in figure~\ref{Figure20a}(b), where also the two local
one-particle levels of the isolated $f$-shell and the spectrum of
the Free Theory are shown. In the latter, two gaps are recognized as
counterparts of those in figure~\ref{Figure20a}(a); the $f$-states
acquire dispersion through mixture with the band and share, in
corresponding regions, the effect of level-repulsion and
gap-formation. 

Interestingly, the CA1-impurity solver at the low
temperature considered, i.e.\ $\beta=15$, is able to describe the
formation of the gap in the narrow region of quasiparticle states
near $\omega=0$, which was to be expected as a consequence of
coherence in the Anderson-lattice. This effect is connected with a
strong increase of scattering at $\omega=0$, see
figure~\ref{Figure20a}(c) where imaginary parts of the "local" self
energies
\begin{eqnarray}
\widetilde{\Sigma}_{a\sigma}(z)=z-\epsilon_a-\Big(\frac{1}{N}\sum_
{\underline{k}}G_{aa\underline{k}\sigma}(z)\Big)^{-1}
\quad,\quad(z=\hbar\omega+i\delta\:,\:a=c,f) \label{Gleichung33}
\end{eqnarray}
are shown; the corresponding very narrow spikes at $\omega=0$ can
cause numerical problems. In approximate impurity solvers usually
convergence problems of the XNCA/DMFT-cycle are observed. Whereas
the original gaps become smeared also in the $f$-spectrum, the
coherence gap should become perfect in the limit $T\rightarrow0$.
This leads to the narrow gapped quasiparticle band structure, to be
seen in the $\underline{k}$-resolved spectra of figure~\ref{Figure20b}(a) 
and (b), and to the broad structures, smeared out by the interactions, at
higher excitation energies. 

Finally it has to be mentioned that
stability of the Fermi-liquid phase might not be thermodynamically stable
for all parameter values used in our
calculations for the Hubbard- and the Anderson-lattice model.
Susceptibilities can point to instabilities towards other possible
ground states~\cite{Quelle60}, which is among a variety of methods
being applied to the investigation of phase diagrams.
\section{Conclusion and outlook}
The foregoing sections have demonstrated the considerable progress,
which has been made in the development of impurity solvers via
direct perturbation theory and their application to impurity- and
lattice-problems with strongly correlated electrons. Our
presentation has emphasized an unified view on several
approximations of this kind, which have been proposed in the past,
and on a new one, the CA1, discussed here for the first time. 

All of
these approximations can be characterized in a systematic fashion as
skeleton expansions in terms of time-ordered local perturbational
processes along the lines laid out by~\cite{Quelle31}. As such, they
furnish coupled implicit integral equations for propagators, which
in general have to be solved numerically; this gives rise to the
notion "semianalytic". ENCA, SUNCA, FNCA and CA1 include
different classes of vertex corrections; the first three of these
approximations reduce to SNCA, the version of the old NCA without
any vertex correction applied to finite $U$. CA1, on the other hand,
contains fully crossing vertex corrections of fourth order in the
hybridization. 

It turns out that for the quality of the
approximation it is important to include ladders for repeated
particle scattering and higher oder vertex corrections in a
well-balanced way. This is apparently accomplished best by the CA1,
which however does not iterate special subclasses to infinite order
like SUNCA and, more generally, FNCA. Comparison with
NRG-calculations in the spirit of Wilsons approach, reveals that
even CA1 has deficiencies at low temperatures and excitation
energies. At higher energies, however, the situations is reversed: In
an overall view taking into account the complete spectral region,
the semianalytical impurity solvers, with the possible exception of
SNCA, perform quite well, and even the ENCA, as the least
complicated of them, may be used for qualitative investigations.

The unified view developed here also concerns the construction of
approximations for lattice problems and the use of the impurity
solvers therein. It was shown that appropriate choices of local
building blocks, each containing a set of internally correlated
one-particle states, and a selection of paths for propagation
through the lattice can be consistently combined in a
matrix-formulation for a calculation of partition function and
Greensfunctions. 

It was explained how in certain simple situations,
as e.g.\ encountered in the Hubbard model or the Anderson-lattice
model with noninteracting band states, the formalism reduces to a
scalar one and how a locally complete selections of local processes
then leads to the well known XNCA-and DMFT-approximations. 

Our
starting point was a cumulant expansion for all local n-particle
vertices, which renders the application of the linked cluster
theorem and of unrestricted site summations possible, thus enabling
a convenient $\underline{k}$-space representation of quantities.
Generalized dynamical fields have been introduced and traced back to
matrix-propagators along closed loops. The neglect of all cumulant
vertices of order $n\geq{2}$ leaves as a natural "Free Theory" the
matrix-generalization of the Hubbard-I approximation, which can
readily be evaluated in explicit form. Restricted selection of
cumulant vertices to finite order allows to define e.g.\ Hartree-type
approximations~\cite{Quelle52,Quelle54}, which are not yet well
investigated but supposedly are of restricted usefulness for the
regime of low temperatures and excitation energies. 

The locally
complete approximations, on the other hand, combine advantages of
real space as well as of $\underline{k}$-space formulations with a
better treatment of infrared divergencies through infinite order.
Nowadays this can be implemented by a variety of local impurity
solvers, among which we have concentrated here on the class based on
direct perturbation theory. As applications of the formalism we have
presented local and $\underline{k}$-resolved one-particle-excitation
spectra for Hubbard- and Anderson-lattice model and have discussed
characteristic similarities and differences. It has proven useful to
connect this discussion with the foregoing treatment of the SIAM as
the prototypical effective impurity. In particular, the formation of
a Fermi liquid could be illuminated in this way, emphazising a local
point of view.

\end{fmfshrink}
\end{fmffile}
\begin{fmffile}{fmf_nca3}
\begin{fmfshrink}{0.8}
\fmfcmd{%
    style_def wiggly_arrow expr p =
     cdraw (wiggly p);
     cfill (arrow p)
    enddef;
    style_def dbl_wiggly_arrow expr p =
     draw_double (wiggly p) ;
     cfill (arrow p);
    enddef;
}
\begin{figure}[th]
  \begin{center}
    \input{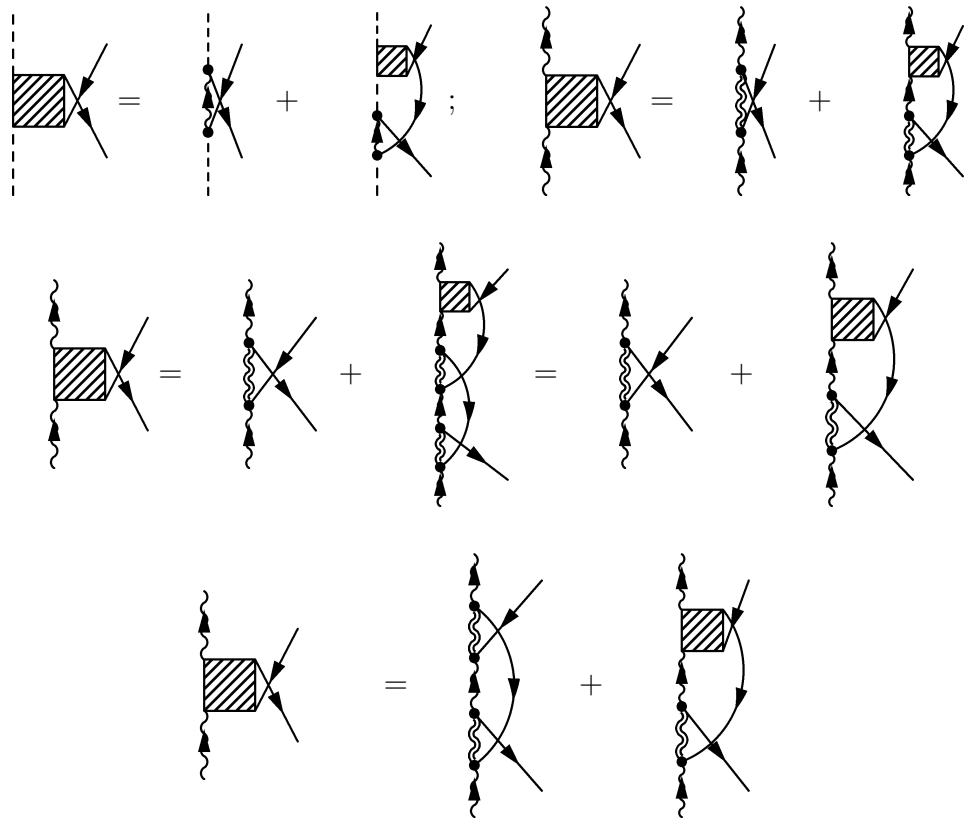}
  \end{center}
  \caption{Higher order corrections to CA1 can be implemented via the four T-matrix 
    equations shown graphically.   The first of these was considered in figure~\ref{Figure5}. 
    The other three equations would generalize
    $U=\infty$-theories like PNCA and CTMA to finite values of U (CA2-project).}
  \label{Figure22}
\end{figure}
\end{fmfshrink}
\end{fmffile}
We will conclude with a short perspective on possible future
developments on the basis of our local approach and with some
critical remarks about its shortcomings. Improvements of
semianalytical impurity solvers could be based on CA1 and proceed
along directions laid down in the simpler case of the
$U=\infty$-version of SIAM~\cite{Quelle41,Quelle44}. 

Two different
approaches could be combined to develop such a CA2-theory. The fully
crossing fourth order vertex corrections could be reinforced by
certain diagrams of sixth order like in figure~\ref{Figure4}(b), and
possibly iterated further, which had proven as beneficial for
spectral properties at large $U$ in the frame of the
PNCA~\cite{Quelle41}. The CTMA~\cite{Quelle44,Quelle46}, on the
other hand as the second of these theories for $U=\infty$, stresses
the role of long ladders of crossing particle lines. 

A
generalization to finite $U$ can be accomplished by solving the
system of four coupled $T$-matrix equations shown in
figure~\ref{Figure22}. The resulting T-matrices then are to be inserted
into the fully crossing vertex corrections contained in CA1; they
replace parts shown as lowest order contributions envisaged in
figure~\ref{Figure3}. The first T-matrix of figure~\ref{Figure22} e.g.\
replaces the diagram in
figure~\ref{Figure4}(a), leading to the sequence of
figure~\ref{Figure5}(b). Both measures together should again be well
balanced in the sense discussed above. Although we expect further
essential quantitative improvements by such additions to CA1, the
resulting CA2-impurity solver can still not be expected to be
perfect. 

It lies at the heart of the infrared problems in SIAM or
its generalizations that no treatment based on a restricted
selection of perturbational processes to infinite order can
adequately describe the complete energy range down to $\omega=0$.
From a practical point of view, however, for many purposes this will
not really be necessary, for example when other types of
correlations in concentrated systems intervene, e.g.\ producing
magnetic states. One problem will be left anyway, even in this case:
The numerical expense connected with approximations like CA1, and
even more so with a hypothetical CA2, is considerable. Certainly
there will be a need for improved algorithms or for the use of
parallel computing. At the end, these higher semianalytical impurity
solvers may turn out impractical compared with e.g.\
Quantum-Monte-Carlo or NRG-methods. Their ability, on the other
hand, to describe the regime at higher temperatures and excitation
energies very well, is already shared by the ENCA, which does not
need so much numerical effort.

Impurity solvers in general serve as a key ingredient in the local
approach to lattice systems. Current research aims at the inclusion
of more realistic local building blocks containing several orbitals
or clusters of sites and at a better treatment of nonlocal
correlations. With an optimal selection of a localized basis of
Wannier-states screened direct interactions are hopefully short
ranged and can either be treated within a small cluster as a local
building block or via an extension of the perturbation in the
Hamiltonian to double-transfer between neighbours, as indicated in
section 2. 

This, as well as a multi-orbital situation, can in
principle be handled by the matrix-propagator formalism presented in
this paper; it has been demonstrated in section 4 how matrix-loop
propagators systematically serve to define appropriate dynamical
fields acting on effective sites. Long ranged correlations, however,
need an extension of the locally complete approximation for their
proper incorporation. 

The cumulant approach outlined above, can
serve as a basis for such generalizations. The nonlocal processes
additionally to be taken into account in e.g.\ selfenergies or
susceptibilities connect local cumulants with four external legs or
more on different sites via more than one loop. The resulting
correlations between propagating particles or holes and between the
states of local blocks on different sites may generate different
types of long-range order, e.g.\ phases of itinerant or local-moment
magnetism. 

Technically, the difficulty in setting up such
appropriate nonlocal theories seems to lie in a consistent choice of
diagram classes as skeletons without overcounting, so that basic
requirements are fulfilled, such as conservation laws and analytic
properties like positivity of excitation spectra. This requires,
apart from a strict use of cumulant subtractions, the consequential
implementation of the concept of irreducibility~\cite{Quelle21}.
Hopefully, work along these lines will soon lead to improved forms
of e.g.\ band structure theories for correlated electrons and will
shed more light on the mechanisms behind the formation of exotic
ground states in transition metal compounds.

\ack
One of the authors (N.G.) expresses his gratitude to the Max Planck
Institut f\"{u}r Physik Komplexer Systeme in Dresden and to its director
Prof. P. Fulde for their hospitality and the opportunity for
discussions and extensive numerical calculations contributing to
this work.

This research was supported in parts (FBA) by the DFG  project AN
275/5-1. FBA also acknowledges supercomputer support by the NIC,
Forschungszentrum  J\"ulich under project no.\ HHB000. 

\section*{References}


\begin{thebibliography}{99}
\bibitem{Quelle1}  Keiter H and  Kimball J C 1971 \textit{Int. J. Magn. {\bf 1}, 233}\\
 Keiter H and  Kimball J C  1971 J. Appl. Phys. {\bf 42}, 1460
\bibitem{Quelle2}  Bringer A and Lustfeld H 1977 \textit{Z. Phys. B {\bf 28}, 213}\\
 Lustfeld H and  Bringer A 1978 \textit{Solid State Commun.  {\bf 28}, 119}
\bibitem{Quelle3}  Grewe  N and  Keiter H 1981 \textit{Phys. Rev. B {\bf 24}, 4420}
\bibitem{Quelle4}  Grewe  N 1982 \textit{Valence Instabilities},
  ed Wachter P and  Boppart H  (North-Holland Publ. Co.)  p~21.
\bibitem{Quelle5} Early review article are:\\
  G\"{u}ntherodt G 1976 \textit{Configurations of 4f Electrons in Rare Earth
    Compounds} "Festk\"{o}rperprobleme XVI / Advances in Solid State
    Physics" (Vieweg \& Sohn, Braunschweig) p~95\\
    Wohlleben D 1976 \textit{J. de Physique Coll.{\bf C4}, 231}
\bibitem{Quelle6} Grewe N and Steglich F 1991 \textit{Heavy Fermions,
    in Handbook on the Physics and Chemistry of Rare Earths, vol.14,}
  ed Gschneidner K A and Eyring L  (Elsevier Science Publ. B.V.)
\bibitem{Quelle7} Steglich F, Aarts J, Bredl C D , Lieke W,
  Meschede D, Franz W and  Sch\"{a}fer H 1979
  \textit{Phys. Rev. {\bf 43}, 1892}
\bibitem{Quelle8} Grewe N 1983 \textit{Z. Physik B-Condensed Matter {\bf 52}, 193  and {\bf 53}, 271 }
\bibitem{Quelle9} Kuramoto Y 1983 \textit{Z. Physik B-Condensed Matter {\bf 53}, 37  and {\bf 54}, 293
    (1984)}
\bibitem{Quelle10} Wilson K G 1975 \textit{Rev. Mod. Phys. 47, 773
    }\\
Krishnamurthy H R ,  Wilkins J W and Wilson K G 1980 \textit{Phys.
  Rev. B{\bf 21}, 1003}
\bibitem{Quelle11} For an early review see:\\
  Gr\"{u}ner  G and  Zawadowski A 1974 \textit{Rep. Progr. Phys. {\bf 37},
  1497}
\bibitem{Quelle12} Baym G and Kadanoff L P 1961 \textit{Phys. Rev. {\bf 124}, 287
    }\\
 Baym G 1962 \textit{Phys. Rev. {\bf 127}, 1391}
\bibitem{Quelle13}  Coleman P 1983 \textit{Phys. Rev. B {\bf 28},
    5255}\\
 Coleman P 1984 \textit{Phys. Rev. B {\bf 29}, 3035 }
\bibitem{Quelle14}  Kuramoto Y and Kojima H 1984 \textit{Z. Physik B-Condensed Matter {\bf 57}, 95}
\bibitem{Quelle15} Kuramoto Y and  M\"{u}ller-Hartmann E 1985 \textit{J. Magnetism and Magn. Materials {\bf 52}, 122}
\bibitem{Quelle16}  Nozi\`{e}res P and  De Dominicis C T 1969
  \textit{Phys. Rev. {\bf 178}, 1097}\\
   Nozi\`{e}res P and  De Dominicis C T 1969 \textit{Phys. Rev. {\bf 178}, 1084 } \\
  Nozi\`{e}res P and  De Dominicis C T 1969 \textit{Phys. Rev. {\bf 178}, 1097 } 
\bibitem{Quelle17}  Menge B and  M\"{u}ller-Hartmann E 1988 \textit{Z. Physik B-Condensed Matter {\bf 73}, 225 }
\bibitem{Quelle18} For a review see:\\
  Cox D L and  Zawadowski A 1998 \textit{Advances in Physics
  {\bf 47}, 599}
\bibitem{Quelle19}  Grewe N 1984 \textit{Solid State Commun. {\bf 50}, 19}
\bibitem{Quelle20}  Kuramoto Y 1985 \textit{Theory of Heavy Fermions and Valence
    Fluctuations}  ed Kasuya T and  Saso T (Springer-Verlag)  p~152
\\
Kim C I,  Kuramoto Y and  Kasuya T 1990 \textit{J. Phys. Soc. Japan
  {\bf 59}, 2414}
\bibitem{Quelle21} Grewe N 1987 \textit{Z. Physik B-Condensed Matter {\bf 67}, 323}\\
 Grewe N,  Pruschke T and  Keiter  H 1988 \textit{Z. Physik B-Condensed
  Matter {\bf 71}, 75}
\bibitem{Quelle22}  Georges A and  Kotliar G 1996 \textit{Rev. Mod. Phys. {\bf 68}, 13}
\bibitem{Quelle23} Brito J J S  and Frota H O 1990 \textit{Phys. Rev. B {\bf 42}, 6378}\\
Costi T A and Hewson A C 1990 \textit{Physica B {\bf 163}, 179 }\\
Costi T A and Hewson A C 1997 \textit{ Phil. Mag. B {\bf 65}, 1165 }
\bibitem{Quelle24}  Bulla R, Costi T and  Pruschke T 2007 \textit{Rev. Mod. Phys.}
\bibitem{Quelle25} Pollmann F,  Runge E and  Fulde P 2006 \textit{Phys. Rev. B {\bf 73}, 125121}
\bibitem{Quelle26}  Schumann R 2002 \textit{Ann. Phys. (Leipzig) {\bf 1}, 49}
\bibitem{Quelle27}  Maier T,  Jarrel M,  Pruschke T and  Hettler M H 2005 \textit{Rev. Mod. Phys.
    {\bf 77}, 1027}
\bibitem{Quelle28}  Held K, Nekrasov I A,  Bl\"{u}mer N,
  Anisimiov V I and Vollhardt D 2001
  \textit{Int. J. Mod. Physics B {\bf 15}, 2611}\\
 Kotliar G and  Vollhardt D 3/2004 \textit{Physics Today, 53}
\bibitem{Quelle29} Grewe N and  Pruschke T 1985 \textit{Z. Physik B-Condensed Matter {\bf 60}, 311}
\bibitem{Quelle30}  Metzner W and  Vollhardt D 1989 \textit{Phys. Rev. Lett. {\bf 62}, 324}
\bibitem{Quelle31}  Keiter H and  Morandi G 1984 \textit{Phys. Rep. {\bf 109}, 227 }
\bibitem{Quelle32} Bickers N E, Cox D L and Wilkins J W 1987 \textit{Phys. Rev. B{\bf 36}, 2036}\\
Bickers N E 1987  \textit{Rev. Mod. Phys. {\bf 59}, 845 }
\bibitem{Quelle33} Kondo J 1964 \textit{Progr. Theor. Physics {\bf 32}, 37}
\bibitem{Quelle34} Ramakrishnan T V 1981 \textit{Valence Fluctuations in
    Solids} 
ed  Falicov L M,  Hanke W  (Maple North-Holland,
  Amsterdam) p~13\\
Ramakrishnan T V and  Sur K 1982 \textit{Phys. Rev. B{\bf 26}, 1798}
\bibitem{Quelle35} Keiter H 1982 \textit{Z. Physik B-Condensed Matter {\bf 49}, 209}
\bibitem{Quelle36}  Nozieres P 1974 \textit{J. Low Temp. Physics {\bf 17}, 31}
\bibitem{Quelle37}  Pruschke T and  Grewe N 1989 \textit{Z. Physik B-Condensed Matter {\bf 74}, 439}
\bibitem{Quelle38}  Keiter H 1985 \textit{Z. Physik B-Condensed Matter {\bf 60}, 337
    }\\
 Keiter H and  Qin Q 1990 \textit{Z. Physik B-Condensed Matter
  {\bf 79}, 397}
\bibitem{Quelle39}  Schrieffer J R  and  Wolff P A 1966 \textit{Phys. Rev {\bf 149}, 491}
\bibitem{Quelle40}  M\"{u}hlschlegel B 1968 \textit{Z. Physik {\bf 208}, 94   }\\
 Coqblin B and  Schrieffer J R 1969 \textit{Phys, Rev. {\bf 185},
  847}
\bibitem{Quelle41} Anders F B and  Grewe N 1994 \textit{Europhys. Lett. {\bf 26}, 551
    }\\
 Anders F B 1995 \textit{J. Phys. Condens. Matter {\bf 7}, 2801}
\bibitem{Quelle42}  Grunenberg J and  Keiter H 1991 \textit{Physica B {\bf 171}, 39}
\bibitem{Quelle43}  Haule K, Kirchner S, Kroha J and W\"{o}lfle P 2001 \textit{Phys. Rev. B {\bf 64}, 155111}
\bibitem{Quelle44} Kroha J,  W\"{o}lfle P and  Costi T A 1997 \textit{Phys. Rev. Lett. {\bf 79}, 261}
\bibitem{Quelle45} Mahan G D 1967 \textit{Phys. Rev {\bf 153}, 882
  }\\
Mahan G D 1967 \textit{Phys. Rev {\bf 163}, 612 }
\bibitem{Quelle46}  Kroha J and  W\"{o}lfle P 2005 \textit{J. Phys. Soc. Japan {\bf 74}, 16 }
\bibitem{Quelle46a} Sakai O, Shimizu1 Y and  Kaneta Y 2005 \textit{J. Phys. Soc. Jpn. {\bf 74}, 2517}\\
  Sakai O, Motizuki M and Kasuya T 1988 in \textit{Core-Level Spectroscopy in Condensed Systems Theory} 
  ed. Kanamori J (Springer, Berlin) p~45\\
   Kang K and  Min B I  1996 \textit{Phys. Rev. B {\bf 54}, 1645}
\bibitem{Quelle46b} Otsuki J and  Kuramoto Y 2006 \textit{J. Phys. Soc. Jpn. {\bf 75}, 064707}
\bibitem{Quelle47}  Anderson P W 1947 \textit{Phys. Rev. Lett. {\bf
      18}, 1049 }\\
 Anderson P W 1967 \textit{Phys. Rev. Lett.  Phys. Rev. {\bf 164}, 352}
\bibitem{Quelle48} Schotte K D and  Schotte U 1969 \textit{Phys. Rev. {\bf 182}, 479}
\bibitem{Quelle49}  Anders F B and  Pruschke T 2006 \textit{Phys. Rev. Lett. {\bf 96}, 086404 }
\bibitem{Quelle50}  Hewson A C 1993 \textit{The Kondo Problem to Heavy
    Fermions} (Cambridge Univ. Press)  p~63
\bibitem{Quelle51}  Hubbard J 1963 \textit{Proc. Royal Soc. A {\bf 276}, 238 }
\bibitem{Quelle52}  Grewe N 1998 \textit{lecture notes ``Lokale Theorie" }
\bibitem{Quelle53}  Grewe N 2005 \textit{Ann. Phys. (Leipzig) {\bf 14}, 611 }\\
  Vladimir M I and  Moskalenko V A 1990 \textit{Theor. Math. Phys. {\bf 82}, 301} 
\bibitem{Quelle54}  Sherman A 2006 \textit{Phys.Rev. B {\bf 73},
    155105}\\
 Sherman A 2006 \textit{Phys.Rev. B {\bf 74}, 035104 }\\
  Vakaru S I, Vladimir M I and  Moskalenko V A 1990 \textit{Theor. Math. Phys. {\bf 85}, 1185      }
\bibitem{Quelle55}  Craco L and  Gusm\~ao M A 1995 \textit{Phys. Rev B {\bf 52}, 17135}\\
 Craco L and  Gusm\~ao M A 1996 \textit{Phys. Rev B {\bf 54}, 1629}\\
 Consiglio R and  Gusm\~ao  M A 1997 \textit{Phys. Rev B {\bf 55}, 6825 }
\bibitem{Quelle56}  Friedel J 1952 \textit{Phil. Mag. {\bf 43}, 153}\\
   Yamada K 1974 \textit{Prog. Theor. Phys. {\bf 53}, 970;}\\
 Yamada K 1975 \textit{Prog. Theor. Phys. {\bf 54}, 316 }
\bibitem{Quelle57} Martin R M and Allen J W 1979 \textit{J. Applied Phys. {\bf 50}, 7561  }\\
  Martin R M 1982 \textit{Phys. Rev. Lett. {\bf 48}, 362}
\bibitem{Quelle57a} 
   Rubtsov A N,  Katsnelson M I and  Lichtenstein A I 2008 \textit{Phys. Rev. B {\bf 77}, 033101}
\bibitem{Quelle58} 
  Jabben T,  Grewe N and  Anders F B 2005 \textit{Eur. Phys. J. {\bf B44}, 47}
\bibitem{QuellePAM}
   Jarrell M 1995 \textit{Phys. Rev B {\bf 51}, 7429 } \\
   Pruschke T,  Bulla R and  Jarrell M 2000 \textit{Phys. Rev B {\bf 61}, 12799 } \\
   Grenzebach C, Anders F B,  Czycholl G and  Pruschke T 2006 \textit{Phys. Rev B {\bf 74}, 195119} 
\bibitem{QuelleHM}
   Jarrell M 1992 \textit{Phys. Rev Lett {\bf 69}, 168} \\
   Pruschke T,  Jarrell M and  Freericks J K 1995 \textit{Adv. Phys. {\bf 44}, 187} 
\bibitem{Quelle59}  Schmitt S and  Grewe N 2005 \textit{Physica {\bf B359-361}, 777 }
\bibitem{Quelle60}  Schmitt S and  Grewe N, \textit{to be published}
\end{thebibliography}
\end{document}